\documentclass[acmtocl,acmnow]{acmtrans2m}

\usepackage{latexsym}
\usepackage{times}
\usepackage{amsmath}
\usepackage[dvips]{graphicx}
\usepackage{mathrsfs}

\newcommand{\mycaption}[2]{\caption[#1]{{\bf #1} #2}}
\newcommand{\comment}[1]{}

\newtheorem{theorem}{Theorem}[section]

\newtheorem{corollary}[theorem]{Corollary}
\newtheorem{proposition}[theorem]{Proposition}
\newtheorem{lemma}[theorem]{Lemma}
\newdef{definition}[theorem]{Definition}
\newdef{remark}[theorem]{Remark}

\newcounter{eg-counter}

\newcommand{\bnf}{\mathrel{::=}}
\newcommand{\kw}[1]{{\bf #1}}
\newcommand{\va}[1]{{\rm {\it #1}}}
\newcommand{\mva}[1]{\mbox{\va{#1}}}
\newcommand{\syntaxform}[1]{\mva{#1}}

\newcommand{\truthe}{\syntaxform{bool-expr}}

\newcommand{\terme}{\syntaxform{int-expr}}

\newcommand{\termc}{\syntaxform{int-constant}}
\newcommand{\functione}{\syntaxform{function-expr}}
\newcommand{\predicatee}{\syntaxform{predicate-expr}}

\newcommand{\termv}{\syntaxform{lambda-var}}
\newcommand{\truths}{\syntaxform{bool-symbol}}
\newcommand{\termsymbol}{\syntaxform{int-symbol}}
\newcommand{\functions}{\syntaxform{function-symbol}}
\newcommand{\predicates}{\syntaxform{predicate-symbol}}

\newcommand{\ITE}{\mva{ITE}}
\newcommand{\ITEe}[3]{\ITE(#1,\,#2,\,#3)}

\newcommand{\true}{\kw{true}}
\newcommand{\false}{\kw{false}}

\newcommand{\compare}[2]{#1\!=\!#2} 
\newcommand{\ncompare}[2]{#1\!\not =\!#2} 
\newcommand{\lessthan}[2]{#1\!<\!#2} 
\newcommand{\incr}[1]{#1\!+\!1}
\newcommand{\decr}[1]{#1\!-\!1}
\newcommand{\vbar}{\mathbin{|}}
\newcommand{\lvbar}{|\;}
\newcommand{\lambdaexpr}[2]{\lambda\, #1 \mathbin{.} #2}

\newcommand{\snameset}[1]{\boldsymbol{#1}}
\newcommand{\nsphi}{\snameset{\phi}}
\newcommand{\nspi}{\snameset{\pi}}

\newcommand{\makesymbol}[1]{\mathtt{#1}}
\newcommand{\makesubsymbol}[1]{\mathtt{#1}}

\newcommand{\makesymbolset}[1]{{\cal #1}}

\newcommand{\statesymbolset}{\makesymbolset{V}}
\newcommand{\inputsymbolset}{\makesymbolset{I}}
\newcommand{\initsymbolset}{\makesymbolset{J}}

\newcommand{\indexsymbolset}{\makesymbolset{X}}
\newcommand{\predsymbolset}{\makesymbolset{P}}
\newcommand{\axiomsymbolset}{\makesymbolset{Q}}

\newcommand{\predeval}[2]{\evalexpr{\nsphi}{\combineinterps{#1}{#2}}}
\newcommand{\predevalsubs}[3]{\evalexpr{\substituteone{\nsphi}{\nssubs}{\indexsymbolset}}{\combineinterps{\combineinterps{#1}{#2}}{#3}}}
\newcommand{\rdepth}{{\it depth}}
\newcommand{\predmap}[1]{#1^{*}}

\newcommand{\myimplies}{\Rightarrow}
\renewcommand{\iff}{\Leftrightarrow}
\newcommand{\pwrset}{{\mathscr P}}

\newcommand{\absstateset}{S_A}
\newcommand{\altabsstateset}{T_A}
\newcommand{\concstateset}{S_C}
\newcommand{\altconcstateset}{T_C}
\newcommand{\concreach}{R_C}
\newcommand{\concinit}{Q^{0}_C}
\newcommand{\absreach}{R_A}
\newcommand{\absinit}{Q^{0}_A}
\newcommand{\absreachform}{\rho_{A}}
\newcommand{\concnext}{N_{C}}
\newcommand{\absnext}{N_{A}}

\newcommand{\sFF}{\mbox{FF}}
\newcommand{\sFT}{\mbox{FT}}
\newcommand{\sTF}{\mbox{TF}}
\newcommand{\sTT}{\mbox{TT}}

\newcommand{\symbolseta}{\makesymbolset{A}}
\newcommand{\symbolsetb}{\makesymbolset{B}}
\newcommand{\symbolsetc}{\makesymbolset{C}}

\newcommand{\exprset}[1]{\ensuremath{E(#1)}}

\newcommand{\nextfunction}{\ensuremath{\delta}}

\newcommand{\indexstate}[1]{q^{#1}}

\newcommand{\substituteone}[3]{\ensuremath{#1 \left[ #2 / #3 \right]}}

\newcommand{\interp}[1]{\ensuremath{\sigma_{\mbox{}\!#1}}}
\newcommand{\interpset}[1]{\ensuremath{\Sigma_{#1}}}
\newcommand{\altinterp}[1]{\ensuremath{\tau_{\mbox{}\!#1}}}

\newcommand{\combineinterps}[2]{\ensuremath{#1 \cdot #2}}

\newcommand{\aabsstateset}{\interpset{\predsymbolset}^{\axiomsymbolset}}

\newcommand{\subs}{\ensuremath{\pi}}
\newcommand{\nssubs}{\ensuremath{\nspi}}
\newcommand{\subsset}{\ensuremath{\Pi}}
\newcommand{\subssetp}{\ensuremath{\Pi'}}

\newcommand{\absreachapprox}{R_{\subsset}}
\newcommand{\absformapprox}{\rho_{\subsset}}
\newcommand{\absreachapproxp}{R_{\subssetp}}
\newcommand{\absapproxnext}{N_{\subsset}}

\newcommand{\evalexpr}[2]{\ensuremath{\left\langle#1\right\rangle_{#2}}}
\newcommand{\evalset}[1]{\ensuremath{\left\langle#1\right\rangle}}

\newcommand{\tobool}[1]{\tilde{#1}}

\newcommand{\LogicName}{CLU}
\newcommand{\FullLogicName}{{Counter arithmetic with Lambda expressions
  and Uninterpreted functions}}

\newcommand{\uclid}{{\sc UCLID}}

\title{Predicate Abstraction with Indexed Predicates}

\author{
SHUVENDU K. LAHIRI\\
RANDAL E. BRYANT\\
Carnegie Mellon University
}

\begin{abstract}
Predicate abstraction provides a powerful tool for verifying
properties of infinite-state systems using a combination of a decision
procedure for a subset of first-order logic and
symbolic methods originally developed for finite-state model checking.
We consider models containing first-order state
variables, where the system state includes mutable functions and
predicates.  Such a model can describe systems containing arbitrarily
large memories, buffers, and arrays of identical processes.  We
describe a form of predicate abstraction that constructs a
formula over a set of universally quantified variables to describe
invariant properties of the first-order state variables.  We provide a formal
justification of the soundness of our approach and describe how it has
been used to verify several hardware and software designs, including
a directory-based cache coherence protocol.
\end{abstract}


\begin{document}

\begin{bottomstuff} 
A shorter version of the paper appeared at {\it International
Conference on Verification, Model Checking and Abstract Interpretation,
(VMCAI) 2004}

\end{bottomstuff}

\maketitle

\section{Introduction}

Graf and Sa\"{\i}di introduced {\em predicate abstraction}
\cite{graf-cav97} as a means of automatically determining invariant
properties of infinite-state systems.  With this approach, the user
provides a set of $k$ Boolean formulas describing possible
properties of the system state.  These predicates are used to generate
a finite state abstraction (containing at most $2^k$ states) of the
system.  By performing a reachability analysis of this finite-state
model, a predicate abstraction tool can generate the strongest
possible invariant for the system expressible in terms of this set of
predicates.  Prior implementations of predicate abstraction
\cite{graf-cav97,saidi-cav99,das-cav99,das-lics01,ball-pldi01,flanagan-popl02,chaki-icse03}
required making a large number of calls to
a theorem prover or first-order decision procedure, and hence could
only be applied to cases where the number of predicates was small.
More recently, we have shown that both BDD and SAT-based Boolean
methods can be applied to perform the analysis efficiently
\cite{lahiri-cav03a}.

In most formulations of predicate abstraction, the predicates contain
no free variables, and hence they evaluate to true or false for each
system state.  The abstraction function $\alpha$ has a
simple form, mapping each {\em concrete} system state to a single
{\em abstract} state based on the effect of evaluating the $k$ predicates.
The task of predicate abstraction is to construct a formula
$\predmap{\psi}$ consisting of some Boolean combination of the
predicates such that $\predmap{\psi}(s)$ holds for every reachable
system state $s$.

To verify systems containing unbounded resources, such as buffers and
memories of arbitrary size and systems with arbitrary numbers of
identical, concurrent processes, the system model must support {\em
first-order} state variables, in which the state variables are
themselves functions or predicates \cite{ip-cav96,bryant-cav02}.  For
example, a memory can be represented as a function mapping an address
to the data stored at an address, while a buffer can be represented as
a function mapping an integer index to the value stored at the
specified buffer position.  The state elements of a set of identical
processes can be modeled as functions mapping an integer process
identifier to the state element for the specified process.  In many
systems, this capability is restricted to {\em arrays} that can be
altered only by writing to a single location
\cite{burch-cav94,mcmillan-cav98}.  Our verifier allows a more general
form of mutable function, where the updating operation is expressed
using lambda notation.

In verifying systems with first-order state variables, we require
quantified predicates to describe global properties of state variables,
such as ``At most one process is in its critical section,'' as
expressed by the formula $\forall i, j : \makesymbol{crit}(i) \land
\makesymbol{crit}(j) \myimplies i = j$.  Conventional predicate
abstraction restricts the scope of a quantifier to within an
individual predicate.  System invariants often involve complex
formulas with widely scoped quantifiers.  The scoping restriction
(the fact that the universal quantifier does not distribute over
conjunctions) implies that these invariants cannot be divided into 
small, simple predicates.  This puts a heavy burden on the user to 
supply predicates that encode intricate sets of properties about the 
system.  Recent work attempts to discover quantified predicates 
automatically \cite{das-fmcad02}, but this is a formidable task.

In this paper we present an extension of predicate abstraction in
which the predicates include free variables
from a set of {\em index} variables $\indexsymbolset$ (and hence the
name {\it indexed predicates}).  The predicate
abstraction engine constructs a formula $\predmap{\psi}$ consisting of
a Boolean combination of these predicates, such that the formula
$\forall \indexsymbolset \predmap{\psi}(s)$ holds for every reachable system
state $s$.  With this method, the predicates can be very simple, with
the predicate abstraction tool constructing complex, quantified
invariant formulas.  For example, the property that at most one
process can be in its critical section could be derived by supplying
predicates $\makesymbol{crit}(\makesymbol{i})$,
$\makesymbol{crit}(\makesymbol{j})$, and
$\makesymbol{i}=\makesymbol{j}$, where $\makesymbol{i}$ and
$\makesymbol{j}$ are the index symbols.  Encoding these predicates in
the abstract system with Boolean variables $\makesymbol{ci}$,
$\makesymbol{cj}$, and $\makesymbol{eij}$, respectively, we can verify
this property by using predicate abstraction to prove that
$\makesymbol{ci} \land \makesymbol{cj} \myimplies \makesymbol{eij}$
holds for every reachable state of the abstract system.

Flanagan and Qadeer use a method similar to ours
\cite{flanagan-popl02}, and we briefly described our method in an
earlier paper \cite{lahiri-cav03a}.  
Our contribution in this paper is
to describe the method more carefully, explore its properties, and
to provide a formal argument for its soundness.
The key idea of our approach is to formulate the abstraction function
$\alpha$ to map a concrete system state $s$ to the set of all possible
valuations of the predicates, considering the set of possible values
for the index variables $\indexsymbolset$.  The resulting abstract
system is unusual; it is not characterized by a state transition
relation and hence cannot be viewed as a state transition system.
Nonetheless, it provides an abstraction interpretation of the
concrete system \cite{cousot-popl77} and hence can be used to find
invariant system properties.

Assuming a decision procedure that can determine the satisfiability of
a formula with universal quantifiers, we prove the following
completeness result: Predicate abstraction can prove any property that
can be proved by induction on the state sequence using an induction
hypothesis expressed as a universally quantified formula over the
given set of predicates.  For many modeling logics, this decision
problem is undecidable.  By using quantifier instantiation, we can
implement a sound, but incomplete verifier.
As an extension, we show that it is easy to incorporate {\em axioms}
into the system, properties that must hold universally for every
system state.  Axioms can be viewed simply as indexed predicates
that must evaluate to true on every step. 

The ideas have been implemented in \uclid{}~\cite{bryant-cav02},
a platform for modeling and verifying infinite-state systems. 
Although we demonstrate the ideas in the context of this tool 
and the logic (\LogicName{}) it supports, the ideas developed 
here are not strongly tied to this logic.
We conclude the paper by describing our
use of predicate abstraction to verify several hardware and software
systems, including a directory-based cache coherence protocol devised
by Steven German \cite{german-cache}.  
We believe we are the
first to verify the protocol for a system with an unbounded number of clients,
each communicating via unbounded FIFO channels.

\subsection{Related Work}
\label{sec:related}
\comment{
Techniques for infinite-state verification.
\begin{itemize}
\item Predicate Abstraction --- initial formulation, work of Graf-Saidi, 
Saidi-Shankar, Slam, Blast, Magic, Das-Dill-Park. 
Quantified predicates used by Das, Lakhnekh. Mention why successive
approximation does not work.
\item Flanagan and Qadeer -- software, sequential programs, we formalise
it. Lakhnekh gives a formulation with free variables. 
\item Invisible Invariants -- automatic, need stratification. 
\item Cut-off based approaches. 
\item Compositional model checking based approach. 
\item Regular model checking
\item Model Checking with network invariants.
\end{itemize}
} 

Verifying systems with unbounded resources is in general undecidable.
For instance, the problem of verifying if a system of $N$ ($N$ can be 
arbitrarily large) concurrent processes satisfies a property is 
undecidable~\cite{apt-ipl86}. Despite its complexity, the problem of
verifying systems with arbitrary large resources (e.g. parameterized
systems with $N$ processes, out-of-order processors with arbitrary large
reorder buffers, software programs with arbitrary large arrays)
is of significant practical interest. Hence, in recent years, 
there has been a lot of interest in developing techniques based
on  model checking and deductive approaches for verifying such systems. 

McMillan uses ``compositional model checking''~\cite{mcmillan-cav98} 
with various built-in abstractions to reduce an infinite-state system 
to a finite state system, which can be model checked using Boolean methods. 
The abstraction mechanisms include
{\it temporal case splitting}, {\it datatype reduction}~\cite{clarke-popl92}
and {\it symmetry reduction}.
Temporal case splitting  uses heuristics to slice the program space to 
only consider the resources necessary for proving a property. 
Datatype reduction uses abstract interpretation~\cite{cousot-popl77}
to abstract unbounded data and operations 
over them to operations over finite domains. For such finite domains,
datatype reduction is subsumed by predicate abstraction. 
Symmetry is exploited to reduce the number of indices to consider 
for verifying unbounded arrays or network of processes. 
The method can prove both safety and liveness properties. 
Since the abstraction mechanisms are built into the system, they can 
often be coarse and may not suffice for proving a system. 
Besides, the user is often required to 
provide auxiliary lemmas or to decompose the proof to be discharged by 
symbolic model checkers. For instance, the proof of safety of the 
Bakery protocol~\cite{mcmillan-cav00} or the proof of an out-of-order processor
model~\cite{mcmillan-cav98} required non-trivial lemmas in the compositional 
model checking framework. 

{\it Regular model checking}~\cite{kesten-cav97,bouajjani-cav00} 
uses regular
languages to represent parameterized systems and computes the
closure for the regular relations to construct the reachable state space.
In general, the method is not guaranteed to be complete and requires various
{\it acceleration} techniques (sometimes guided by the user) to ensure 
termination. 
Moreover, approaches based on regular language are not suited for 
representing data in the system. Several examples that we consider in this
work can't be modeled in this framework; the 
out-of-order processor which contains data operations 
or the Peterson's mutual exclusion
are few such examples. Even though the Bakery algorithm can
be verified in this framework, it requires considerable user ingenuity to 
encode the protocol in a regular language. 


Several researchers have investigated restrictions on the system 
description to make the parameterized verification problem decidable. 
Notable among them
is the early work by German and Sistla~\cite{german-jacm92} for verifying
single-indexed properties for synchronously communicating systems.
For restricted systems, finite ``cut-off'' based 
approaches~\cite{emerson-popl95,emerson-cade00,emerson-charme03a} 
reduce the problem to verifying networks of some fixed finite size. 
These bounds have been established for verifying restricted classes of
ring networks and cache coherence protocols. 
Emerson and Kahlon~\cite{emerson-charme03a} have verified the version
of German's cache coherence protocol with single entry channels by 
manually reducing it to a snoopy protocol, for which finite cut-off
exists. 
However, the reduction is manually performed and exploits details of operation
of the protocol, and thus requires user ingenuity.
It can't be easily extended to verify other unbounded systems including the 
Bakery algorithm or the out-of-order processors.

The method of ``invisible invariants''~\cite{pnueli-tacas01,arons-cav01}
uses heuristics for constructing universally quantified invariants 
for parameterized systems automatically. The method computes the set 
of reachable states for finite (and small) instances of the parameters 
and then generalizes them to parameterized systems to construct a
potential inductive invariant. They provide an algorithm for checking the 
verification conditions for a restricted class of system called the
{\it stratified} systems, which include German's protocol with single
entry channels and Lamport's Bakery protocol~\cite{lamport-74}.
However, the method simply becomes a heuristic for generating 
candidate invariants for non-stratified systems, which includes
Peterson's mutual exclusion algorithm~\cite{peterson-ipl81} and 
the Ad-hoc On-demand Distance Vector (AODV)~\cite{aodv} network
protocol. The class of {\it bounded-data} systems (where each variable
is finite but parameterized) considered by this approach can't model the 
the out-of-order processor model~\cite{lahiri-fmcad02} 
that we can verify using our method.

Predicate abstraction with locally quantified 
predicates~\cite{das-fmcad02,baukus-vmcai02} require complex quantified 
predicates to construct the inductive assertions, as mentioned in the
introduction. These predicates are often as complex as invariants 
themselves. In fact, some of the invariants are used are predicates 
in~\cite{baukus-vmcai02} to derive inductive invariants.
The method in~\cite{baukus-vmcai02} verified (both safety
and liveness) a
version of the cache coherence protocol with single entry channels, 
with complex manually provided predicates.
Baukus et al.~\cite{baukus-vmcai02} uses the the logic of {\it WSIS} 
(weak second order logic with one successor)~\cite{buchi-60,thomas-90},
which does not allow function symbols and thus can't model the 
out-of-order processor model. 
The automatic predicate discovery methods for quantified 
predicates~\cite{das-fmcad02} have not been demonstrated on 
most examples (except the AODV model) we consider in this paper.

Flanagan and Qadeer~\cite{flanagan-popl02} use indexed predicates
to synthesize loop invariants for sequential software programs 
that involve unbounded arrays. They also provide heuristics to 
extract some of the predicates from the program text automatically. 
The heuristics are specific to loops in sequential software and not suited for 
verifying more general unbounded systems that we handle in this 
paper. In this work, we explore formal properties of this formulation
and apply it for verifying distributed systems. 
In a recent work~\cite{lahiri-cav04a},
we provide a weakest precondition transformer~\cite{dijstra-cacm75} based
syntactic heuristic for discovering most of the predicates for many
of the systems that we consider in this paper. 

\section{Notation}
\label{sec:notation}
Rather than using the common {\em indexed vector} notation to represent
collections of values (e.g., $\vec{v} \doteq \langle v_1, v_2, \ldots,
v_n \rangle$), we use a {\em named set} notation.  That is, for a set of
symbols $\symbolseta$, we let $v$ indicate a set
consisting of a value $v_{\makesubsymbol{x}}$ for each $\makesymbol{x}
\in \symbolseta$.

For a set of symbols $\symbolseta$, let
$\interp{\symbolseta}$ denote an {\em interpretation} of these
symbols, assigning to each symbol $\makesymbol{x} \in
\symbolseta$ a value
$\interp{\symbolseta}(\makesymbol{x})$ of the appropriate type
(Boolean, integer, function, or predicate).  Let
$\interpset{\symbolseta}$ denote the set of all interpretations
$\interp{\symbolseta}$ over the symbol set $\symbolseta$.

For interpretations $\interp{\symbolseta}$ and $\interp{\symbolsetb}$
over disjoint symbol sets $\symbolseta$ and $\symbolsetb$, let
$\combineinterps{\interp{\symbolseta}}{\interp{\symbolsetb}}$ denote
an interpretation assigning either
$\interp{\symbolseta}(\makesymbol{x})$ or
$\interp{\symbolsetb}(\makesymbol{x})$ to each symbol $\makesymbol{x}
\in \symbolseta \cup \symbolsetb$, according to whether
$\makesymbol{x} \in \symbolseta$ or $\makesymbol{x} \in \symbolsetb$.

Figure~\ref{fig:syntax} displays the syntax of the Logic of 
\FullLogicName{} (\LogicName{}), a fragment of first-order logic 
extended with equality, inequality, and counters. An {\it expression} 
in  \LogicName{} can evaluate to truth values ($\truthe$), integers 
($\terme$), functions ($\functione$) or predicates ($\predicatee$).
Notice that we only allow restricted arithmetic on terms, namely that 
of addition or subtraction by constants. 
Notice that we restrict the parameters to a lambda expression to be
integers, and not function or predicate expressions.  There is
no way in our logic to express any form of iteration or recursion.

\begin{figure}[h]
\begin{eqnarray*}
\truthe & \bnf  & \true \vbar \false \vbar \truths \\
&& \lvbar \neg \truthe 
 \vbar (\truthe \land \truthe) \\
&& \lvbar  (\compare{\terme}{\terme})  
 \vbar  (\lessthan{\terme}{\terme})  \\
&& \lvbar \predicatee(\terme, \ldots, \terme) \\
\terme & \bnf  & \termv \vbar \termsymbol \\
&& \lvbar \ITEe{\truthe}{\terme}{\terme} \\
&& \lvbar \terme + \termc \\
&& \lvbar \functione(\terme, \ldots, \terme) \\
\predicatee & \bnf & \predicates \vbar \lambdaexpr{\termv, \ldots, \termv}{\truthe} \\
\functione & \bnf & \functions \vbar \lambdaexpr{\termv, \ldots, \termv}{\terme} 
\end{eqnarray*}
\mycaption{\LogicName{} Expression Syntax.}{Expressions can denote 
computations of Boolean values, integers, or functions yielding 
Boolean values or integers.}
\label{fig:syntax}
\end{figure}

For symbol set $\symbolseta$, let $\exprset{\symbolseta}$ denote the
set of all \LogicName{} expressions over $\symbolseta$.  For any expression
$\phi \in \exprset{\symbolseta}$ and interpretation
$\interp{\symbolseta} \in \interpset{\symbolseta}$, let the {\em
valuation of $\phi$ with respect to $\interp{\symbolseta}$}, denoted
$\evalexpr{\phi}{\interp{\symbolseta}}$ be the (Boolean, integer, 
function, or predicate) value obtained by evaluating $\phi$ when each symbol
$\makesymbol{x} \in \symbolseta$ is replaced by its interpretation
$\interp{\symbolseta}(\makesymbol{x})$.

Let ${v}$ be a named set over symbols $\symbolseta$, consisting of
expressions over symbol set $\symbolsetb$.
That is,
$v_{\makesubsymbol{x}} \in \exprset{\symbolsetb}$ for each
$\makesymbol{x} \in \symbolseta$.  Given an interpretation
$\interp{\symbolsetb}$ of the symbols in $\symbolsetb$, evaluating the
expressions in ${v}$ defines an interpretation of the symbols in
$\symbolseta$, which we denote $\evalexpr{{v}}{\interp{\symbolsetb}}$.
That is, $\evalexpr{{v}}{\interp{\symbolsetb}}$ is an interpretation
$\interp{\symbolseta}$ such that $\interp{\symbolseta}(\makesymbol{x})
= \evalexpr{v_{\makesubsymbol{x}}}{\interp{\symbolsetb}}$ for each
$\makesymbol{x} \in \symbolseta$.

A {\em substitution} $\subs$ for a set of symbols $\symbolseta$ is a
named set of expressions over some set of symbols $\symbolsetb$ (with
no restriction on the relation between $\symbolseta$ and
$\symbolsetb$.)  That is, for each $\makesymbol{x} \in \symbolseta$,
there is an expression $\subs_{\makesubsymbol{x}} \in
\exprset{\symbolsetb}$.  For an expression $\psi \in
\exprset{\symbolseta \cup \symbolsetc}$, we let
$\substituteone{\psi}{\subs}{\symbolseta}$ denote the expression
$\psi' \in \exprset{\symbolsetb \cup \symbolsetc}$ resulting when we
replace each occurrence of each symbol $\makesymbol{x} \in
\symbolseta$ with the expression $\subs_{\makesubsymbol{x}}$.  These
replacements are all performed simultaneously.

\begin{proposition}
Let $\psi$ be an expression in $\exprset{\symbolseta \cup \symbolsetc}$ and
$\subs$ be a substitution having $\subs_{\makesubsymbol{x}} \in
\exprset{\symbolsetb}$ for each $\makesymbol{x} \in \symbolseta$.
For interpretations $\interp{\symbolsetb}$ and $\interp{\symbolsetc}$, if we let $\interp{\symbolseta}$ be the interpretation defined as
$\interp{\symbolseta} =
\evalexpr{\pi}{\interp{\symbolsetb}}$, then
$\evalexpr{\psi}{\combineinterps{\interp{\symbolseta}}{\interp{\symbolsetc}}} =
\evalexpr{\substituteone{\psi}{\subs}{\symbolseta}}{\combineinterps{\interp{\symbolsetb}}{\interp{\symbolsetc}}}$.
\label{prop:eval-sub}
\end{proposition}
This proposition captures a fundamental relation between syntactic
substitution and expression evaluation, sometimes referred to as {\em
referential transparency}.  We can interchangeably use a subexpression
$\pi_{\makesubsymbol{x}}$ or the result of evaluating this subexpression
$\interp{\symbolseta}(\makesymbol{x})$ in evaluating a formula
containing this subexpression.

\section{System Model}

We model the system as having a number of {\em state elements}, where
each state element may be a Boolean or integer value, or a function or
predicate.  We use symbolic names to represent the different state
elements giving the set of {\em state symbols} $\statesymbolset$.  We
introduce a set of {\em initial state} symbols $\initsymbolset$ and a set of
 {\em input} symbols $\inputsymbolset$
representing, respectively, initial values and
inputs that can be set to arbitrary
values on each step of operation.  Among the state variables, there can be
{\em immutable} values expressing the behavior of functional
units, such as ALUs, and system parameters such as the total number of
processes or the maximum size of a buffer.  Since these values are
expressed symbolically, one run of the verifier can prove
the correctness of the system for arbitrary functionalities, process
counts, and buffer capacities.

The overall system operation is characterized by an {\em initial-state} expression set $\indexstate{0}$ and a {\em next-state}
expression set $\nextfunction$.  The initial state consists of an
expression for each state element, with the initial value of state element
$\makesymbol{x}$ given by expression
$\indexstate{0}_{\makesubsymbol{x}} \in \exprset{\initsymbolset}$.  The transition behavior also consists of an
expression for each state element, with the behavior for state element
$\makesymbol{x}$ given by expression
$\nextfunction_{\makesubsymbol{x}} \in \exprset{\statesymbolset \cup
\inputsymbolset}$.  In this expression, the state element
symbols represent the current system state and the input symbols 
represent the current values of the inputs.  The expression gives
the new value for that state element.

We will use a very simple system as a running example throughout this
presentation.  The only state element is a function $\makesymbol{F}$,
i.e. $\statesymbolset$ = \{$\makesymbol{F}$\}. 
An input symbol $\makesymbol{i}$ determines which 
element of $\makesymbol{F}$ is updated.  Initially, $\makesymbol{F}$ is the
identify function: 
\begin{equation*}
\indexstate{0}_{\makesubsymbol{F}} = \lambdaexpr{u}{u}.
\end{equation*}  

On each step, the value of the function for
argument $\makesymbol{i}$ is updated to be
$\makesymbol{F}(\incr{\makesymbol{i}})$.  That is,
\begin{equation*}
\nextfunction_{\makesubsymbol{F}} =
\lambdaexpr{u}{\ITEe{u = \makesymbol{i}}{\makesymbol{F}(\incr{\makesymbol{i}})}{\makesymbol{F}(u)}}
\end{equation*}
where the if-then-else operation $\ITE$ selects its second argument
when the first one evaluates to true and the third otherwise.
For the above example, $\initsymbolset = \{ \}$ and 
$\inputsymbolset$ = \{$\makesymbol{i}$\}. 

\comment{
Thus, we have $\statesymbolset = \{ \makesymbol{F} \}$,
$\initsymbolset = \emptyset$, and
and $\inputsymbolset = \{ \makesymbol{i} \}$.
}

\subsection{Concrete System}

A concrete system state assigns an interpretation to every state
symbol.  The set of states of the concrete system is given by
$\interpset{\statesymbolset}$, the
set of interpretations of the state element symbols.  
For convenience, we denote concrete states using
letters $s$ and $t$ rather than the more formal
$\interp{\statesymbolset}$.

From our system model, we can characterize the behavior of the
concrete system in terms of an initial state set 
$\concinit \subseteq \interpset{\statesymbolset}$ and a
next-state function operating on sets $\concnext\colon 
\pwrset({\interpset{\statesymbolset}}) \rightarrow 
\pwrset({\interpset{\statesymbolset}})$.
The initial state set is defined as:
$$
\concinit \doteq \{\evalexpr{\indexstate{0}}{\interp{\initsymbolset}}|\interp{\initsymbolset} \in \interpset{\initsymbolset}\},
$$
i.e.,
the set of all possible valuations of the initial state expressions.
The next-state function $\concnext$ is defined for a single state $s$ as:
$$
\concnext(s) \doteq \{\evalexpr{\nextfunction}{\combineinterps{s}{\interp{\inputsymbolset}}}|
\interp{\inputsymbolset} \in \interpset{\inputsymbolset}\},
$$
i.e., the set of all valuations of the next-state
expressions for concrete state $s$ and arbitrary input.
The function is then extended to sets of states by
defining 
$$\concnext(\concstateset) = \bigcup_{s \in \concstateset}
\concnext(s).$$  We can also characterize the next-state behavior of
the concrete system by a transition relation $T$ where $(s,t) \in T$
when $t \in \concnext({s})$.

We define the set of reachable
states $\concreach$ as containing those states $s$ such that there is
some state sequence $s_0, s_1, \ldots, s_n$ with $s_0 \in \concinit$,
$s_n = s$, and $s_{i+1} \in \concnext(s_i)$ for all values of $i$ such that
$0 \leq i < n$.  We define the {\em depth} of a reachable state $s$ to be
the length $n$ of the shortest sequence leading to $s$.  Since our
concrete system has an infinite number of states, there is no finite
bound on the maximum depth over all reachable states.

With our example system, the concrete state set consists of
integer functions $f$ such that $f(\incr{u}) \geq f(u) \geq u$ for all $u$
and $f(u) = u$ for infinitely many arguments of $f$.

\section{Predicate Abstraction with Indexed Predicates}

We use {\it indexed} predicates to express constraints on the system
state.  To define the abstract state space, we introduce a set of {\em
predicate} symbols $\predsymbolset$ and a set of {\em index} symbols
$\indexsymbolset$.  The predicates consist of a named set $\phi$, where
for each $\makesymbol{p} \in \predsymbolset$, predicate
$\phi_{\makesubsymbol{p}}$ is a Boolean formula over the symbols in
$\statesymbolset \cup \indexsymbolset$.

Our predicates define an abstract state space
$\interpset{\predsymbolset}$, consisting of all interpretations
$\interp{\predsymbolset}$ of the predicate symbols.
For $k \doteq | \predsymbolset |$, 
the state space contains $2^k$ elements.



As an illustration, suppose for our example system we wish to prove
that state element $\makesymbol{F}$ will always be a function $f$ satisfying $f(u) \geq 0$
for all $u    \geq 0$.  We introduce an index variable $\makesymbol{x}$ and
predicate symbols $\predsymbolset = \{ \makesymbol{p}, \makesymbol{q}
\}$, with $\phi_{\makesubsymbol{p}} \doteq
\makesymbol{F}(\makesymbol{x})
\geq 0$ and $\phi_{\makesubsymbol{q}} \doteq \makesymbol{x} \geq 0$.



We can denote a set of abstract states by a Boolean formula $\psi \in
\exprset{\predsymbolset}$.  This expression defines a set of states
$\evalset{\psi} \doteq \{ \interp{\predsymbolset} |
\evalexpr{\psi}{\interp{\predsymbolset}} = \true \}$.  As an example,
our two predicates $\phi_{\makesubsymbol{p}}$ and
$\phi_{\makesubsymbol{q}}$
generate an
abstract space consisting of four elements, which we denote
$\sFF$, $\sFT$, $\sTF$, and $\sTT$, according to the interpretations assigned to 
$\makesymbol{p}$ and $\makesymbol{q}$.
There are then 16
possible abstract state sets, some of which are shown in Table
\ref{tab:eg-concrete}.  In this table, abstract state sets are
represented both by Boolean formulas over $\makesymbol{p}$ and
$\makesymbol{q}$, and by enumerations of the state elements.

\begin{table}
\begin{center}
\begin{tabular}{|c|c|c|c|}
\hline
\multicolumn{2}{|c|}{Abstract System} & \multicolumn{2}{|c|}{Concrete System} \\ 
\hline
\makebox[2.5cm]{Formula} & \makebox[2.5cm]{State Set} & \makebox[3.2cm]{System Property} & \makebox[3.2cm]{State Set} \\
$\psi$ & $\absstateset = \evalset{\psi}$ 
& $\forall \indexsymbolset \predmap{\psi}$
& $\concstateset = \gamma(\absstateset)$ \\
\hline
$\makesymbol{p} \land \makesymbol{q}$ & $\{ \sTT\}$ 
 & $\forall \makesymbol{x} : \makesymbol{f}(\makesymbol{x}) \geq 0 \land \makesymbol{x} \geq 0$
 & $\emptyset$ \\
$\makesymbol{p} \land \neg \makesymbol{q}$ & $\{ \sTF\}$ 
 & $\forall \makesymbol{x} : \makesymbol{f}(\makesymbol{x}) \geq 0 \land \makesymbol{x} < 0$
 & $\emptyset$ \\
$\neg \makesymbol{q}$ & $\{ \sFF, \sTF\}$ 
 & $\forall \makesymbol{x} : \makesymbol{x} < 0$
 & $\emptyset$ \\
$\makesymbol{p}$ & $\{ \sTF, \sTT\}$ 
 & $\forall \makesymbol{x} : \makesymbol{f}(\makesymbol{x}) \geq 0$
 & $\{ f | f(x) \geq 0 \}$ \\
$\makesymbol{p} \lor \neg \makesymbol{q}$ & $\{ \sFF, \sTF, \sTT\}$ 
 & $\forall \makesymbol{x} : \makesymbol{x} \geq 0 \myimplies \makesymbol{f}(\makesymbol{x}) \geq 0$
 & $\{ f | x \geq 0 \myimplies f(x) \geq 0 \}$ \\
\hline
\end{tabular}
\end{center}
\mycaption{Example abstract state sets and their concretizations}{Abstract state elements are represented by their
interpretations of $\makesymbol{p}$ and $\makesymbol{q}$.}
\label{tab:eg-concrete}
\end{table}

We define the {\em abstraction function}
$\alpha$ to map each concrete state to the set of abstract states given by the valuations of the predicates for all possible values of the index variables:
\begin{eqnarray}
\alpha(s) 
& \doteq &
\left \{ \predeval{s}{\interp{\indexsymbolset}} | 
\interp{\indexsymbolset} \in \interpset{\indexsymbolset} \right\}
\label{eqn:single-alpha}
\\
& = & \displaystyle{\bigcup_{\interp{\indexsymbolset} \in \interpset{\indexsymbolset}}
\left \{ \predeval{s}{\interp{\indexsymbolset}} \right\} }
 \label{eqn:alt-single-alpha}
\end{eqnarray}

Note that (\ref{eqn:alt-single-alpha}) is simply a restatement of (\ref{eqn:single-alpha}) using set
union notation.

\begin{figure}[h]
 \centerline{\includegraphics*[scale=0.6]{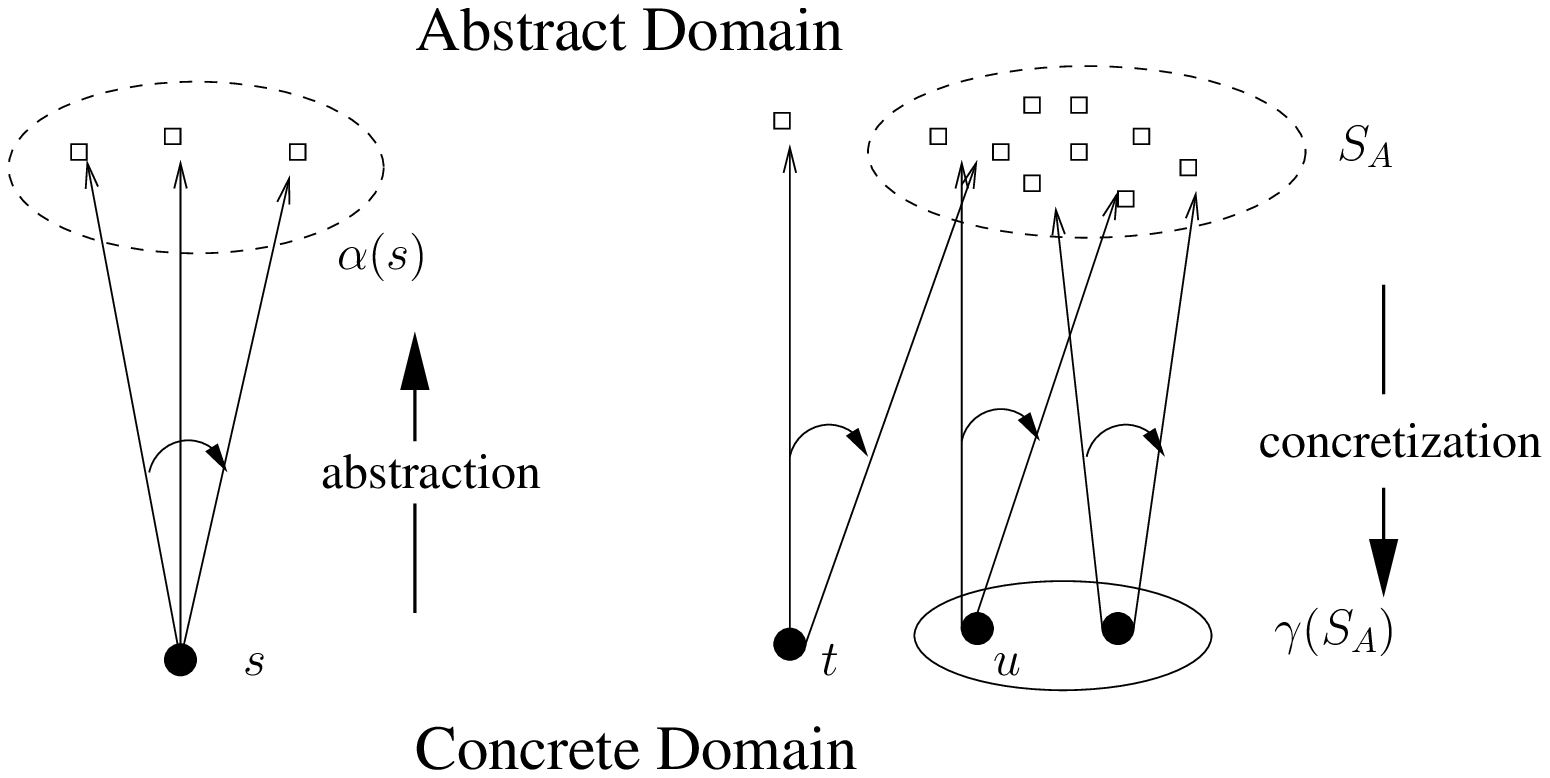}}
\mycaption{Abstraction and Concretization.}{}
\label{fig:abs_conc}
\end{figure}

Since there are multiple interpretations $\interp{\indexsymbolset}$, a
single concrete state will generally map to multiple abstract states.  
Figure~\ref{fig:abs_conc} illustrates this fact. 
The abstraction function $\alpha$ maps a single concrete state $s$  
to a set of abstract 
states --- each abstract state ($\predeval{s}{\interp{\indexsymbolset}}$)
resulting from  some interpretation $\interp{\indexsymbolset}$.  This
feature is not found in most uses of predicate abstraction, but it is
the key idea for handling indexed predicates. 

\begin{figure}[h]
 \centerline{\includegraphics*[scale=0.5]{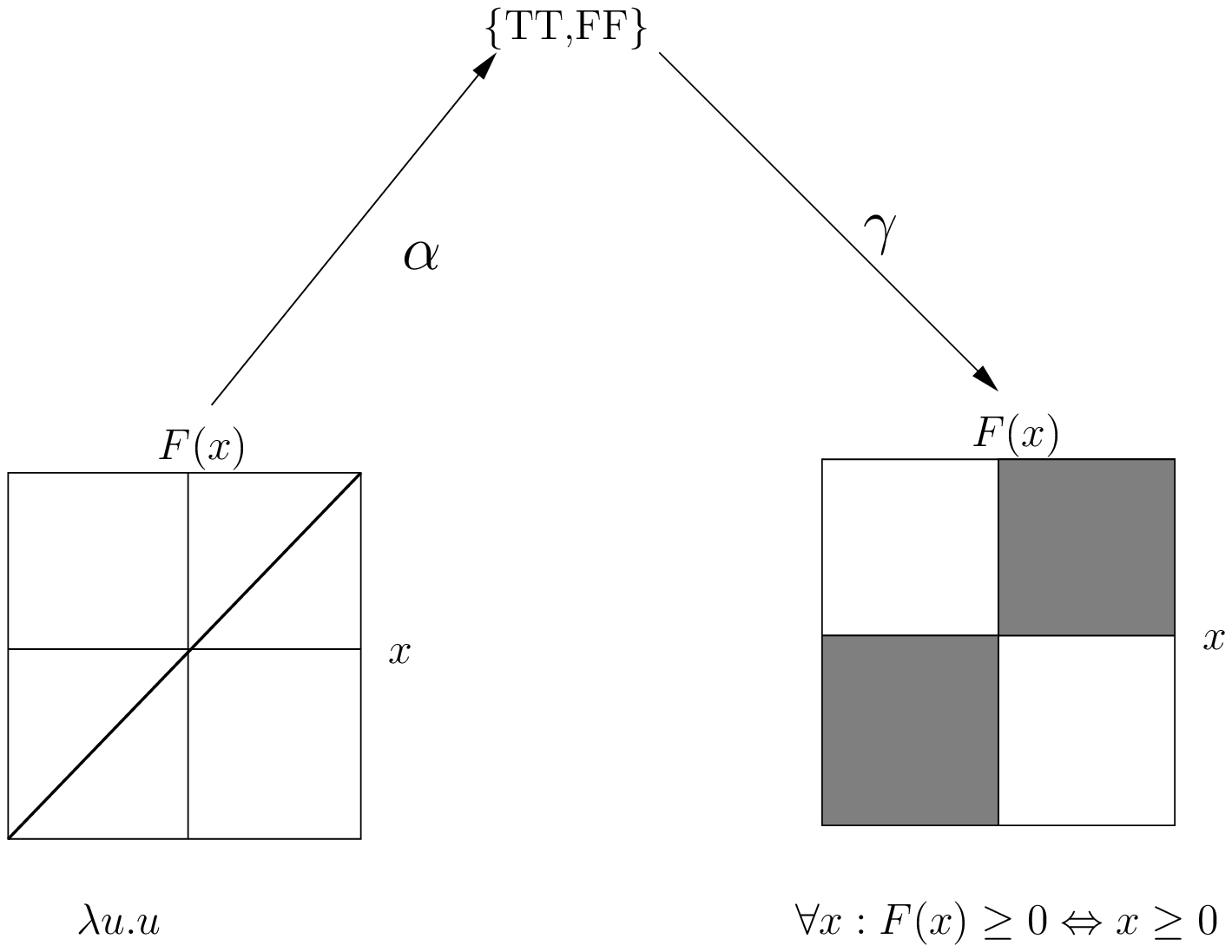}}
\mycaption{Abstraction and Concretization for the initial state for the example.}{}
\label{fig:example-init-abs}
\end{figure}

Working with our example system, consider the concrete state given by the 
function $\lambdaexpr{u}{u}$, in Figure~\ref{fig:example-init-abs}. 
When we abstract this function relative to
predicates $\phi_{\makesubsymbol{p}}$ and 
$\phi_{\makesubsymbol{q}}$, we get two abstract
states: $\sTT$, when $\makesymbol{x} \geq 0$, and $\sFF$, when
$\makesymbol{x} < 0$.  This abstract state set is then characterized
by the formula $\makesymbol{p} \iff \makesymbol{q}$. 

We then extend the abstraction function to apply to sets of concrete states
in the usual way:
\begin{eqnarray}
\alpha(\concstateset) & \doteq & \displaystyle{\bigcup_{s \in \concstateset} \alpha(s).}
\label{eqn:set-alpha}\\
& = & 
\displaystyle{\bigcup_{\interp{\indexsymbolset} \in \interpset{\indexsymbolset}}
\;\;
\bigcup_{s \in \concstateset}
\predeval{s} {\interp{\indexsymbolset}}}
\label{eqn:alt-set-alpha}
\end{eqnarray}

Note that (\ref{eqn:alt-set-alpha}) follows by combining 
(\ref{eqn:alt-single-alpha}) with (\ref{eqn:set-alpha}), and then 
reordering the unions.

\begin{proposition}
For any pair of concrete state sets $\concstateset$ and 
$\altconcstateset$:
\begin{enumerate}
\item If $\concstateset \subseteq \altconcstateset$, then
$\alpha(\concstateset) \subseteq \alpha(\altconcstateset)$.
\item
$\alpha(\concstateset) \cup \alpha(\altconcstateset) =
\alpha(\concstateset \cup \altconcstateset)$.
\end{enumerate}
\label{prop:alpha-properties}
\end{proposition}
These properties follow directly from the way we extended $\alpha$
from a single concrete state to a set of concrete states.

We define the concretization function $\gamma$ to require universal quantification over the index symbols.
That is, for a set of abstract states $\absstateset \subseteq \interpset{\predsymbolset}$, we let
$\gamma(\absstateset)$ be the following set of concrete states:
\begin{equation}
\gamma(\absstateset)
 \;\;\; \doteq \;\;\;
\left\{ s
 | \forall \interp{\indexsymbolset} \in \interpset{\indexsymbolset} :
\predeval{s}
{\interp{\indexsymbolset}} \in \absstateset
\right \}
\label{eqn:gamma}
\end{equation}
Consider the Figure~\ref{fig:abs_conc}, where a set of abstract states
$\absstateset$ has been concretized to a set of concrete states 
$\gamma(\absstateset)$.
It shows a  concrete state $t$ that is not included in 
$\gamma(\absstateset)$ because one of the states it abstracts
to lies outside $\absstateset$. On the other hand, the concrete state 
$u$ is contained in $\gamma(\absstateset)$ because 
$\alpha(u) \subseteq \absstateset$. One can provide an alternate definition
of $\gamma$ as follows:
\begin{equation}
\gamma(\absstateset)
 \;\;\; \doteq \;\;\;
\left\{ s
 | \alpha (s) \subseteq \absstateset
\right \}
\label{eqn:gamma-alt}
\end{equation}

The universal quantifier in the definition of $\gamma$ 
has the consequence that
the concretization function does not distribute over set union.  In
particular, we cannot view the concretization function as operating on
individual abstract states, but rather as generating each
concrete state from multiple abstract states.


\begin{proposition}
For any pair of abstract state sets $\absstateset$ and
$\altabsstateset$:
\begin{enumerate}
\item If $\absstateset \subseteq \altabsstateset$, then
$\gamma(\absstateset) \subseteq \gamma(\altabsstateset)$.
\item
$\gamma(\absstateset) \cup \gamma(\altabsstateset) \subseteq
\gamma(\absstateset \cup \altabsstateset)$.
\end{enumerate}
\label{prop:gamma-properties}
\end{proposition}
The first property follows from (\ref{eqn:gamma}), while the second
follows from the first.

Consider our example system with predicates 
$\phi_{\makesubsymbol{p}}$ and $\phi_{\makesubsymbol{q}}$.
Table \ref{tab:eg-concrete} shows some example abstract state sets
$\absstateset$ and their concretizations $\gamma(\absstateset)$.
As the first three examples show,
some (altogether 6) nonempty abstract state sets have empty concretizations, 
because they constrain
$\makesymbol{x}$ to be either always negative or always nonnegative.
On the other hand, there
are 9 abstract state sets
having nonempty concretizations.  We can see by this that the
concretization function is based on the entire abstract state set and
not just on the individual values.
For example, the sets $\{ \sTF\}$ and $\{\sTT\}$ have empty 
concretizations, but $\{ \sTF, \sTT\}$ concretizes to the set of all 
nonnegative functions.


\begin{theorem}
The functions $(\alpha, \gamma)$ form a Galois connection, i.e., for any sets of concrete states $\concstateset$ and abstract states $\absstateset$:
\begin{eqnarray}
\alpha(\concstateset) \subseteq \absstateset \;\iff\;
\concstateset \subseteq \gamma(\absstateset)
\label{eqn:galois-primary}
\end{eqnarray}
\label{thm:galois}
\end{theorem}

\begin{proof}
(This is one of several logically equivalent formulations of a Galois
connection \cite{cousot-popl77}.)  The
proof follows by observing that both the left and the right-hand sides
of (\ref{eqn:galois-primary}) hold precisely when for every 
$\interp{\indexsymbolset} \in \interpset{\indexsymbolset}$ and every
$s \in \concstateset$, we have 
$\predeval{s}{\interp{\indexsymbolset}} \in \absstateset$. Let us prove
the two directions:
\begin{enumerate}
\item{\it If }: Let $\alpha(\concstateset) \subseteq \absstateset$. By
the definition of $\alpha$ in (\ref{eqn:single-alpha}), this implies 
that for every $s \in \concstateset$ and for interpretation
$\interp{\indexsymbolset} \in \interpset{\indexsymbolset}$, 
$\predeval{s}{\interp{\indexsymbolset}} \in \absstateset$. By the 
definition of $\gamma$ in (\ref{eqn:gamma}), $\gamma(\absstateset)$
contains precisely those concrete states $s'$ for which 
$\predeval{s'}{\interp{\indexsymbolset}} \in \absstateset$, for 
every interpretation 
$\interp{\indexsymbolset} \in \interpset{\indexsymbolset}$. Thus,
for every $s \in \concstateset$, $s \in \gamma(\absstateset)$ and 
consequently, $\concstateset \subseteq \gamma(\absstateset)$.

\item{\it Only if }: Let $\concstateset \subseteq \gamma(\absstateset)$.
Hence, by (\ref{eqn:gamma}), for every $s \in \concstateset$, 
$\predeval{s}{\interp{\indexsymbolset}} \in \absstateset$, for 
every interpretation 
$\interp{\indexsymbolset} \in \interpset{\indexsymbolset}$. By the
definition of $\alpha$ in (\ref{eqn:single-alpha}), 
$\alpha(s) \in \absstateset$. Further, by extending $\alpha$ for the
entire set $\concstateset$ by (\ref{eqn:set-alpha}), 
we get $\alpha(\concstateset) \subseteq \absstateset$.

\end{enumerate}
\end{proof}

Alternately, the functions ($\alpha,\gamma$) form a Galois connection
if they satisfy  the following properties for
any sets of concrete states $\concstateset$ and abstract states 
$\absstateset$:
\begin{eqnarray}
\concstateset \;\; &\subseteq& \;\; \gamma(\alpha(\concstateset)).
\label{eqn:galois-gamma} \\
\alpha(\gamma(\absstateset)) \;\; &\subseteq& \;\; \absstateset.
\label{eqn:galois-alpha} 
\end{eqnarray}
These properties can be derived from (\ref{eqn:galois-primary}). 
Similarly, (\ref{eqn:galois-primary}) can be derived from 
(\ref{eqn:galois-gamma}) and (\ref{eqn:galois-alpha}).
The containment relation in both (\ref{eqn:galois-gamma}) and 
(\ref{eqn:galois-alpha}) can be proper.  
For example, the concrete state set consisting of the single
function $\lambdaexpr{u}{u}$ abstracts to the state set $\makesymbol{p}
\iff \makesymbol{q}$, which in turn concretizes to the set
of all functions $f$ such that $f(u) \geq 0 \iff u \geq 0$, for 
any argument $u$. This is clearly demonstrated in 
Fig~\ref{fig:example-init-abs}.
On the other hand, consider the set of abstract states represented
by $\makesymbol{p} \wedge \makesymbol{q}$. This set of abstract
states has an empty concretization (see Table~\ref{tab:eg-concrete}), 
and thereby satisfies $\alpha(\gamma(\absstateset)) \subset \absstateset$.

\section{Abstract System}

Predicate abstraction involves performing a reachability analysis over
the abstract state space, where on each step we concretize the
abstract state set via $\gamma$, apply the concrete next-state function,
and then abstract the results via $\alpha$.  We can view this process as performing reachability analysis on an abstract system having
initial state set $\absinit \doteq
\alpha(\concinit)$ and a next-state function operating on sets:
$\absnext(\absstateset) \doteq
\alpha(\concnext(\gamma(\absstateset)))$.  Note that there is no
transition relation associated with this next-state function, since
$\gamma$ cannot be viewed as operating on individual abstract states.

It can be seen that $\absnext$ provides an {\em abstract
interpretation}~\cite{cousot-popl77} of the concrete system:
\begin{enumerate}
\item $\absnext$ is null-preserving: $\absnext(\emptyset) = \emptyset$
\item $\absnext$ is monotonic:
$\absstateset \subseteq \altabsstateset\; \myimplies\;
\absnext(\absstateset) \subseteq \absnext(\altabsstateset)$.
\item $\absnext$ simulates $\concnext$ (with a simulation relation defined by $\alpha$):
$\alpha(\concnext(\concstateset)) \subseteq
\absnext(\alpha(\concstateset))$.
\end{enumerate}

\begin{theorem}
$\absnext$ provides an {\em abstract
interpretation} of the concrete transition system $\concnext$.
\label{thm:abs-interp}
\end{theorem}

\begin{proof} 
Let us prove the three properties mentioned above:
\begin{enumerate}
\item This follows from the definition of $\absnext$ and the fact that
$\gamma(\emptyset) = \emptyset$, $\concnext(\emptyset) = \emptyset$ and
$\alpha(\emptyset) = \emptyset$. 

\item By the definition of $\absnext$, and using the fact that 
$\gamma$, $\alpha$ and $\concnext$ are monotonic. $\concnext$ is monotonic
since it distributes over the elements of a set of concrete states,
i.e. $\concnext(\concstateset) = \bigcup_{s \in \concstateset}
\concnext(s)$.

\item From (\ref{eqn:galois-gamma}), we know that $\concstateset \subseteq
\gamma (\alpha(\concstateset))$. By the monotonicity of $\concnext$, 
$\concnext(\concstateset) \subseteq \concnext(\gamma (\alpha(\concstateset)))$.
Since $\alpha$ is monotonic, we have 
$\alpha(\concnext(\concstateset)) \subseteq 
\alpha(\concnext(\gamma (\alpha(\concstateset))))$. Now applying the 
definition of $\absnext$, we get the desired result. 
\end{enumerate}
\end{proof}


\section{Reachability Analysis}

Predicate abstraction involves performing a reachability analysis over
the abstract state space, where on each step we concretize the
abstract states via $\gamma$, apply the concrete transition relation,
and then abstract the results via $\alpha$.  
In particular, define $\absreach^i$, 
the set of states reached on step $i$ as:
\begin{eqnarray}
\absreach^{0} & = & \absinit  \label{eqn:abs-reach-init}
\\
\displaystyle
\absreach^{i+1} & = & 
\absreach^i \cup \absnext(\absreach^i)
\\
& = & 
\displaystyle \absreach^{i} \;\cup \bigcup_{s \in \gamma(\absreach^{i})} \;\;
\bigcup_{t \in \concnext(s)}
\alpha(t)
\label{eqn:abs-reach-next}
\end{eqnarray}

\begin{proposition}
If $s$ is a reachable state in the concrete system such that
$\rdepth(s) \leq n$, then $\alpha(s) \subseteq \absreach^{n}$.
\end{proposition}

\begin{proof}
We prove this by induction on $n$.  For $n = 0$, the only concrete
states having depth $0$ are those in $Q_0$, and by
(\ref{eqn:abs-reach-init}), these states are all included in
$\absreach^0$.

For a state $t$ having depth $k < n$, our induction hypothesis shows
that $\alpha(t) \subseteq \absreach^{n-1}$.  Since $\absreach^{n-1}
\subseteq \absreach^{n}$, we therefore have $\alpha(t) \subseteq
\absreach^{n}$.

Otherwise, suppose state $t$ has depth $n$.  Then there must be some
state $s$ having depth $n-1$ such that $t \in \concnext(s)$.  By the
induction hypothesis, we must have $\alpha(s) \subseteq \absreach^{n-1}$.
By (\ref{eqn:galois-gamma}), we have $s \in \gamma(\alpha(s))$, and
Proposition \ref{prop:gamma-properties} then implies that $s \in
\gamma(\absreach^{n-1})$.  By (\ref{eqn:abs-reach-next}), we can therefore
see that $\alpha(t) \subseteq \absreach^{n}$.
\end{proof}

Since the abstract system is finite, there must be some $n$ such that
$\absreach^{n} = \absreach^{n+1}$.  The set of all reachable abstract
states $\absreach$ is then $\absreach^{n}$.

\begin{proposition}
\label{prop:absapprox}
The abstract system computes an overapproximation of the set of reachable concrete states, i.e.,
\begin{eqnarray}
\alpha(\concreach) & \subseteq & \absreach
\end{eqnarray}
\end{proposition}

Thus, even though determining the set of reachable concrete states
would require examining paths of unbounded length, we can compute a
conservative approximation to this set by performing a bounded reachability
analysis on the abstract system.

\begin{remark}
It is worth noting that we cannot use the standard ``frontier
set'' optimization in our reachability analysis.  This optimization,
commonly used in symbolic model checking, considers only the newly
reached states in computing the next set of reachable states.  In our
context, this would mean using the computation $\absreach^{i+1}  = 
\absreach^i \cup \absnext(\absreach^i-\absreach^{i-1})$ rather than
that of (\ref{eqn:abs-reach-next}).  This optimization is not valid,
due to the fact that $\gamma$, and therefore $\absnext$, does not
distribute over set union.
\end{remark}

\begin{figure}[h]
 \centerline{\includegraphics*[scale=0.5]{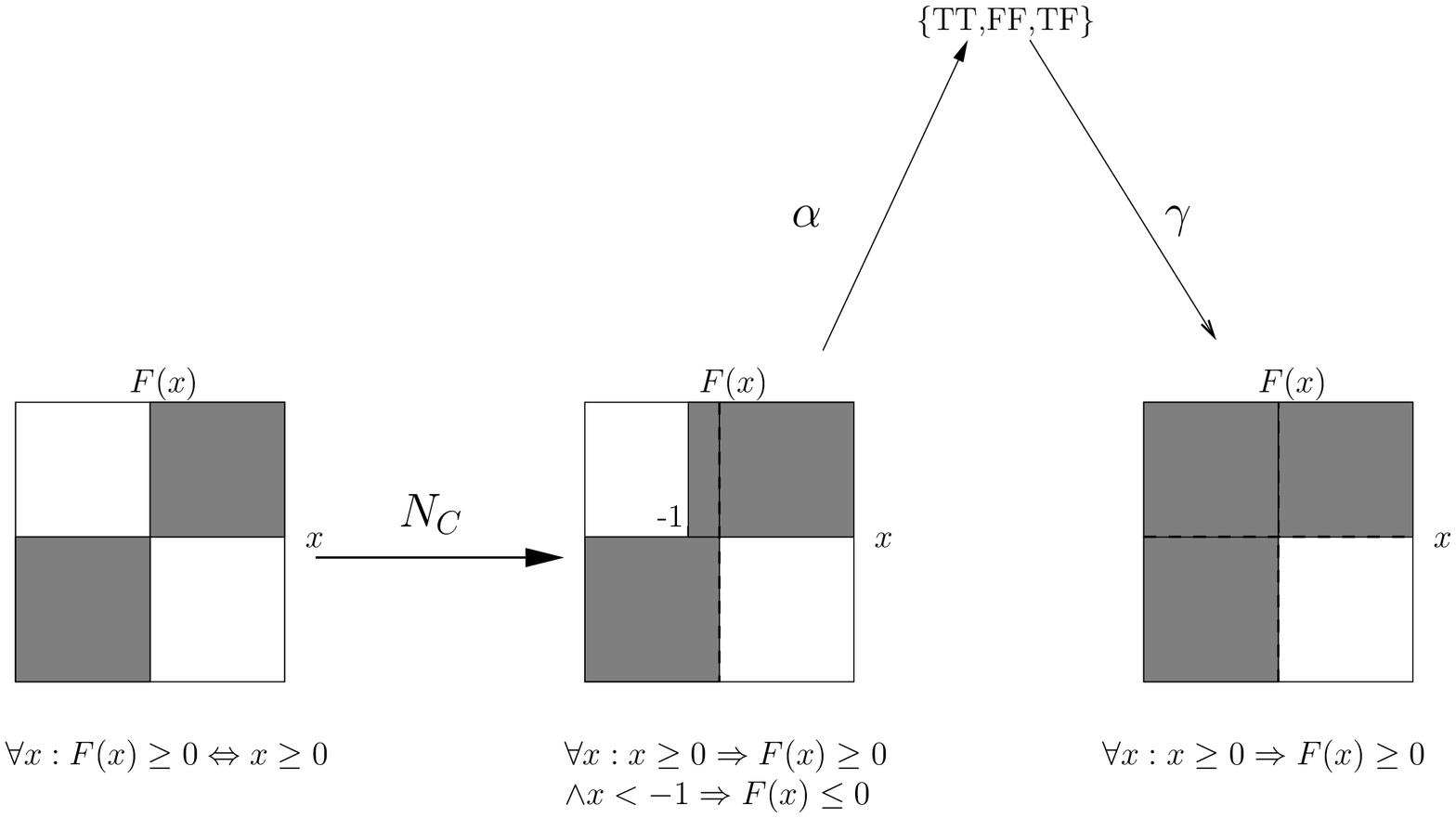}}
\mycaption{Reachability after 1 iteration for the example.}{}
\label{fig:example-step1-reach}
\end{figure}

As an illustration, let us perform reachability analysis on our
example system:
\begin{enumerate}
\item In the initial state, state element $\makesymbol{F}$
is the identity function, which we have seen abstracts to the set
represented by the formula $\makesymbol{p} \iff \makesymbol{q}$.  This
abstract state set concretizes to the set of functions $f$ satisfying 
$f(u) \geq 0 \iff u \geq 0$.  This is illustrated in 
Fig~\ref{fig:example-init-abs}.

\item Let $h$ denote the value of $\makesymbol{F}$ in the next state.
If input $\makesymbol{i}$ is $-1$, we would $h(-1) = f(0) \geq
0$, but we can still guarantee that $h(u) \geq 0$ for $u \geq 0$. 
This is illustrated in Fig~\ref{fig:example-step1-reach}.
Applying the abstraction function, we get 
$\absreach^1$ characterized by the formula $\makesymbol{p} \lor \neg
\makesymbol{q}$ (see Table \ref{tab:eg-concrete}.)   

\item For the second iteration, the abstract state set characterized 
by the formula $\makesymbol{p} \lor \neg \makesymbol{q}$
concretizes to the set of functions $f$ satisfying $f(u) \geq 0$
when $u \geq 0$, and this condition must hold in the next state as well.
Applying the abstraction function to this set, we then get
$\absreach^2 = \absreach^1$, and hence the process has converged.
\end{enumerate}

\section{Verifying Safety Properties}

A Boolean formula $\psi \in \exprset{\predsymbolset}$ can be viewed as
defining a {\em property} of the abstract state space.  Such a
property is said to hold for the abstract system when it holds for
every reachable abstract state.  That is,
$\evalexpr{\psi}{\interp{\predsymbolset}} =
\true$ for all $\interp{\predsymbolset} \in \absreach$.

For Boolean formula $\psi \in \exprset{\predsymbolset}$, define the
formula $\predmap{\psi} \in \exprset{\statesymbolset 
\cup \indexsymbolset}$ to be the result of
substituting the predicate expression $\phi_{\makesubsymbol{p}}$ for each
predicate symbol $\makesymbol{p} \in \predsymbolset$.  That is,
viewing $\phi$ as a substitution, we have $\predmap{\psi} \doteq
\substituteone{\psi}{\phi}{\predsymbolset}$.

\begin{proposition}
For any formula $\psi \in \exprset{\predsymbolset}$, any concrete
 state $s$, and interpretation
$\interp{\indexsymbolset} \in
\interpset{\indexsymbolset}$, if $\interp{\predsymbolset} =
\predeval{s}{
\interp{\indexsymbolset}}$, then
$\evalexpr{\predmap{\psi}}{\combineinterps{s}{\interp{\indexsymbolset}}}
= \evalexpr{\psi}{\interp{\predsymbolset}}$.
\label{prop:abseval}
\end{proposition}
This is a particular instance of Proposition \ref{prop:eval-sub}.

We can view the formula $\predmap{\psi}$ as defining a property
$\forall \indexsymbolset \predmap{\psi}$ of the concrete state space.
This property is said to hold for concrete state $s$, written
$\forall \indexsymbolset \predmap{\psi}(s)$, when
$\evalexpr{\predmap{\psi}}{\combineinterps{s}{\interp{\indexsymbolset}}} = \true$
for every $\interp{\indexsymbolset} \in \interpset{\indexsymbolset}$.
The property is said to hold for the concrete system when $\forall
\indexsymbolset \predmap{\psi}(s)$ holds for every reachable concrete
state $s \in \concreach$.

With our example system, letting formula $\psi \doteq \makesymbol{p}
\lor \neg \makesymbol{q}$, and noting that $\makesymbol{p} \lor \neg
\makesymbol{q} \;\equiv\; \makesymbol{q} \Rightarrow \makesymbol{p}$,
we get as a property of state variable $\makesymbol{F}$ that $\forall
\makesymbol{x} : \makesymbol{x} \geq 0 \Rightarrow
\makesymbol{F}(\makesymbol{x}) \geq 0$.

\begin{proposition}
Property 
$\forall \indexsymbolset \predmap{\psi}(s)$ holds for concrete state $s$
if and only if $\evalexpr{\psi}{\interp{\predsymbolset}} = \true$ for every
$\interp{\predsymbolset} \in \alpha(s)$.
\label{prop:property}
\end{proposition}
This property follows from the definition of $\alpha$ (Equation
\ref{eqn:single-alpha}) and Proposition \ref{prop:abseval}.

Alternately, a Boolean $\psi \in \exprset{\predsymbolset}$ formula can also 
be viewed as characterizing a set of abstract states 
$\evalexpr{\psi}{} \doteq \{\interp{\predsymbolset} \;|\; 
\evalexpr{\psi}{\interp{\predsymbolset}} = \true\}$. Similarly, we can
interpret the formula $\forall \indexsymbolset \predmap{\psi}$
as characterizing the set of concrete states 
$\evalexpr{\forall \indexsymbolset \predmap{\psi}}{} \doteq
\{s\; |\; \evalexpr{\forall \indexsymbolset \predmap{\psi}}{s} = \true\}$.

\begin{proposition}
If $\concstateset \doteq \evalexpr{\forall \indexsymbolset \predmap{\psi}}{}$
and $\absstateset \doteq \evalexpr{\psi}{}$, then $\concstateset = 
\gamma(\absstateset)$.
\label{prop:gamma-formula}
\end{proposition}

\begin{proof}
Expanding the definition of $\concstateset$, we get 
\begin{eqnarray}
\concstateset & =  & \{s \;|\; \forall \interp{\indexsymbolset} \in 
\interpset{\indexsymbolset} \;:\;\evalexpr{\predmap{\psi}}{\combineinterps{s}{\interp{\indexsymbolset}}} = \true \} \label{eqn:gamma-formula-1} 
\\
& =  & \{s \;|\; \forall \interp{\indexsymbolset} \in 
\interpset{\indexsymbolset} \;:\; \interp{\predsymbolset} =
\predeval{s}{\interp{\indexsymbolset}} \; \myimplies \;\evalexpr{\psi}{\interp{\predsymbolset}} = \true \} \label{eqn:gamma-formula-2}
\\ 
& =  & \{s \;|\; \forall \interp{\indexsymbolset} \in 
\interpset{\indexsymbolset} \;:\; 
\predeval{s}{\interp{\indexsymbolset}} \; \in \absstateset \} \label{eqn:gamma-formula-3}
\end{eqnarray}
Observe that (\ref{eqn:gamma-formula-2}) follows from 
(\ref{eqn:gamma-formula-1}) by expanding the definition of $\concstateset$
and (\ref{eqn:gamma-formula-3}) follows from (\ref{eqn:gamma-formula-2})
by using Proposition \ref{prop:abseval}.
\end{proof}

The purpose of predicate abstraction is to provide a way to verify
that a property $\forall \indexsymbolset \predmap{\psi}(s)$ holds for
the concrete system based on the set of reachable abstract states.

\begin{theorem}
For a formula $\psi \in \exprset{\predsymbolset}$, if property $\psi$ holds for the abstract system, then property 
$\forall \indexsymbolset \predmap{\psi}$ holds for the concrete system.
\label{thm:safety}
\end{theorem}

\begin{proof}
Consider an arbitrary concrete state $s \in
\concreach$ and an arbitrary interpretation $\interp{\indexsymbolset}
\in \interpset{\indexsymbolset}$.
If we let $\interp{\predsymbolset} =
\predeval{s}{\interp{\indexsymbolset}}$, then by the definition of 
$\alpha$ (Equation \ref{eqn:single-alpha}), we must have
$\interp{\predsymbolset} \in \alpha(s)$.  By Propositions 
\ref{prop:alpha-properties} and \ref{prop:absapprox}, we therefore have
\begin{displaymath}
\interp{\predsymbolset} 
\;\;\; \in \;\;\; 
 \alpha(s)
 \;\;\; \subseteq \;\;\; 
\alpha(\concreach)
 \;\;\; \subseteq \;\;\; 
\absreach
\end{displaymath}
By the premise of the theorem we have
$\evalexpr{\psi}{\interp{\predsymbolset}} = \true$, and by Proposition
\ref{prop:abseval}, we have
$\evalexpr{\predmap{\psi}}{\combineinterps{s}{\interp{\indexsymbolset}}}
= \evalexpr{\psi}{\interp{\predsymbolset}} = \true$.  This is
precisely the condition required for the
property $\forall \indexsymbolset \predmap{\psi}$ to hold for
 the concrete system.
\end{proof}

Thus, the abstract reachability analysis on our example system does
indeed prove the property that any value $f$ of state variable
$\makesymbol{F}$ satisfies $\forall x : x \geq 0 \Rightarrow f(x) \geq
0$.

Using predicate abstraction, we can possibly get a {\em false
negative} result, where we fail to verify a property $\forall
\indexsymbolset \predmap{\psi}$, even though it holds for the concrete
system, because the given set of predicates does not adequately capture the
characteristics of the system that ensure the desired property.  Thus,
this method of verifying properties is sound, but possibly incomplete.

For example, any reachable state $f$ of
our example system satisfies $\forall x : f(x) < 0 \Rightarrow f(-x)
\geq -x$, but our reachability analysis cannot show this.

We can, however, precisely characterize the class of properties for
which this form of predicate analysis is both sound and complete.
A property $\forall \indexsymbolset \predmap{\psi}$ is said to be {\em inductive} for the concrete system when it satisfies the following two properties:
\begin{enumerate}
\item Every initial state $s \in Q_0$ satisfies $\forall \indexsymbolset \predmap{\psi}(s)$.
\item For every pair of concrete states $(s,t)$, such that 
$t \in \concnext(s)$, if 
$\forall \indexsymbolset \predmap{\psi}(s)$ holds, then so does
$\forall \indexsymbolset \predmap{\psi}(t)$.
\end{enumerate}

\begin{proposition}
If $\forall \indexsymbolset \predmap{\psi}$ is inductive, then 
$\forall \indexsymbolset \predmap{\psi}$ holds for the concrete system. 
\end{proposition}
This proposition follows by induction on the state sequence leading to
each reachable state.

Let $\absreachform$ be a formula that exactly characterizes the set of
reachable abstract states.  That is, $\evalset{\absreachform} = \absreach$.

\begin{lemma}
\label{lemma:absreachform}
$\forall \indexsymbolset
\predmap{\absreachform}$ is inductive.
\end{lemma}

\begin{proof}
By definition, $\evalexpr{\absreachform}{\interp{\predsymbolset}} =
\true$ if and only if $\interp{\predsymbolset} \in \absreach$, and so
by Proposition \ref{prop:property}, $\forall \indexsymbolset
\predmap{\absreachform}(s)$ holds for concrete state $s$ if and only if
$\alpha(s) \subseteq \absreach$.

We can see that the first requirement is satisfied for any $s \in
 Q_0$, since $\alpha(s) \subseteq \alpha(Q_0) \subseteq \absreach$ and
 therefore $\forall \indexsymbolset \predmap{\absreachform}(s)$ holds
by Proposition \ref{prop:property}.

Now suppose there is a state $t \in \concnext(s)$ and $\forall
\indexsymbolset \predmap{\absreachform}(s)$ holds.  Then we must have
$\alpha(s) \subseteq \absreach^{i}$ for some $i \geq 0$.  From
(\ref{eqn:galois-gamma}), we have $s \in \gamma(\alpha(s)) \subseteq
\gamma(\absreach^{i})$, and therefore, by (\ref{eqn:abs-reach-next}), $\alpha(t) \subseteq
\absreach^{i+1} \subseteq \absreach$.  Thus, the second requirement is
satisfied.
\end{proof}

\begin{lemma}
\label{lemma:inductive}
If $\forall \indexsymbolset \predmap{\psi}$ is inductive, then 
$\psi$ holds for the abstract system.
\end{lemma}

\begin{proof}
We will prove by induction on $i$ that
$\evalexpr{\psi}{\interp{\predsymbolset}} = \true$ for every
$\interp{\predsymbolset} \in \absreach^{i}$.  From the definition of
$\absreach$, it then follows that
$\evalexpr{\psi}{\interp{\predsymbolset}} = \true$ for every
$\interp{\predsymbolset} \in \absreach$, and therefore $\psi$ holds for
the abstract system.

For the case of $i = 0$, (\ref{eqn:abs-reach-init}) indicates that
$\absreach^{0} = \alpha(Q_0)$.  Thus, by the definition of $\alpha$
(Equation \ref{eqn:single-alpha}) for every $\interp{\predsymbolset} \in
\absreach^{0}$, there must be a state $s$
and an interpretation $\interp{\indexsymbolset} \in
\interpset{\indexsymbolset}$ such that $\interp{\predsymbolset}
= \predeval{s}{\interp{\indexsymbolset}}$.  
By the first property of an inductive
predicate and by Proposition \ref{prop:abseval}, we have
$\evalexpr{\psi}{\interp{\predsymbolset}} =
\evalexpr{\predmap{\psi}}{\combineinterps{s}{\interp{\indexsymbolset}}}
= \true$.

Now suppose that $\evalexpr{\psi}{\interp{\predsymbolset}} = \true$
for all $\interp{\predsymbolset} \in \absreach^{i}$.
Consider
an element $\altinterp{\predsymbolset} \in \absreach^{i+1}$.  If
$\altinterp{\predsymbolset} \in \absreach^{i}$, then our induction
hypothesis shows that $\evalexpr{\psi}{\altinterp{\predsymbolset}} =
\true$.  Otherwise, by (\ref{eqn:abs-reach-next}), and the definitions
of $\alpha$ (Equation \ref{eqn:single-alpha}), the transition relation 
$\concnext$, and
$\gamma$ (Equation \ref{eqn:gamma}), there must be concrete states $s$
and $t$ satisfying:
\begin{enumerate}
\item $\altinterp{\predsymbolset} \in \alpha(t)$.  That is,
$\altinterp{\predsymbolset}
= \predeval{t}{\altinterp{\indexsymbolset}}$
for some $\altinterp{\indexsymbolset} \in \interpset{\indexsymbolset}$.
\item $t \in \concnext(s)$.
\item $s \in \gamma(\absreach^{i})$.  That is, for all 
$\interp{\indexsymbolset} \in \interpset{\indexsymbolset}$, if
$\interp{\predsymbolset} \doteq \predeval{s}{\interp{\indexsymbolset}}$, then
$\interp{\predsymbolset} \in \absreach^{i}$.
\end{enumerate}
By Proposition \ref{prop:abseval} we have 
$\evalexpr{\predmap{\psi}}{\combineinterps{s}{\interp{\indexsymbolset}}}
= \evalexpr{\psi}{\interp{\predsymbolset}} = \true$, and therefore
$\forall \indexsymbolset \predmap{\psi}(s)$ holds.  By the second property of an inductive predicate, 
$\forall \indexsymbolset \predmap{\psi}(t)$ must also hold.  Applying Proposition \ref{prop:abseval} once again, we therefore have
$\evalexpr{\psi}{\altinterp{\predsymbolset}} = 
\evalexpr{\predmap{\psi}}{\combineinterps{t}{\altinterp{\indexsymbolset}}}
= \true$.  This completes our induction.
\end{proof}

This lemma simply shows that if we present our predicate
abstraction engine with a fully formed induction hypothesis, then it
will be able to perform the induction proof.  But, it has 
important consequences.

For a formula $\psi \in \exprset{\predsymbolset}$ and a predicate set
$\phi$, the property $\forall \indexsymbolset \predmap{\psi}$ is said
to {\em have an induction proof over $\phi$} when there is some
formula $\chi \in \exprset{\predsymbolset}$, such that $\chi
\Rightarrow \psi$ and $\forall \indexsymbolset \predmap{\chi}$ is
inductive.  That is, there is some way to strengthen $\psi$ into a
formula $\chi$ that can be used to prove the property by induction.

\begin{theorem}
A formula $\psi \in \exprset{\predsymbolset}$ is a property of the
abstract system if and only if the concrete property $\forall \indexsymbolset
\predmap{\psi}$ has an induction proof over the predicate set $\phi$.
\label{thm:induction}
\end{theorem}

\begin{proof}
Suppose there is a formula $\chi$ such that $\forall \indexsymbolset
\predmap{\chi}$ is inductive.  Then by Lemma \ref{lemma:inductive}, we
know that $\chi$ holds in the abstract system, and when $\chi
\Rightarrow \psi$, we can infer that $\psi$ holds in the abstract
system.

On the other hand, suppose that $\psi$ holds in the abstract system.
Then the formula $\absreachform$ (characterizing the set of all
reachable abstract states) satisfies $\absreachform \Rightarrow \psi$
and $\forall \indexsymbolset \predmap{\absreachform}$ is inductive.
Hence $\forall \indexsymbolset \predmap{\psi}$ has an induction proof
over $\phi$.
\end{proof}

This theorem precisely characterizes the capability of our formulation
of predicate abstraction --- it can prove any property that can be
strengthened into an induction hypothesis using some combination of
the predicates.  Thus, if we fail to verify a system using this form
of predicate abstraction, we can conclude that either 1) the system
does not satisfy the property, or 2) we did not provide an adequate
set of predicates out of which the predicate abstraction engine could
construct a universally quantified induction hypothesis.

\begin{corollary}
The property $\forall \indexsymbolset \predmap{\absreachform}$ is the
strongest inductive invariant for the concrete system of the form
$\forall \indexsymbolset \predmap{\chi}$, where $\chi \in 
\exprset{\predsymbolset}$. Alternately,  for any other inductive property
$\forall \indexsymbolset \predmap{\chi}$, where $\chi \in 
\exprset{\predsymbolset}$, 
$\forall \indexsymbolset \predmap{\absreachform} \myimplies
\forall \indexsymbolset \predmap{\chi}$. 
\label{cor:strongest-inv}
\end{corollary}
\begin{proof}
The proof follows easily from Theorem \ref{thm:induction}, the fact 
that $\absreachform \Rightarrow \chi$ whenever $\chi$ is a property of the
abstract state space, Proposition \ref{prop:gamma-formula} and 
Proposition~\ref{prop:gamma-properties}. 
\end{proof}

\begin{remark}
To fully automate the process of generating invariants, we need to 
further discover the predicates automatically. Other predicate abstraction
tools~\cite{ball-pldi01,henzinger-popl02,chaki-icse03,das-fmcad02} 
generate new predicates based on ruling out spurious counterexample 
traces from the abstract model.  This approach cannot be used
directly in our context, since our abstract system cannot be viewed as
a state transition system, and so there is no way to characterize a
counterexample by a single state sequence.
In this paper, we do not address the issue of discovering the 
indexed predicates: we provide a syntactic heuristic based on the weakest
precondition transformer in a separate work~\cite{lahiri-cav04a}. 
\end{remark}

\section{Quantifier Instantiation}
For many subsets of first-order logic, there is no complete method for
handling the universal quantifier introduced in function $\gamma$
(Equation \ref{eqn:gamma}).  For example, in a logic with
uninterpreted functions and equality, determining whether a
universally quantified formula is satisfiable is undecidable
\cite{borger-97}.  Instead, we concretize abstract states by
considering some limited subset of the interpretations of the index
symbols, each of which is defined by a substitution for the symbols in
$\indexsymbolset$.  Our tool automatically generates candidate substitutions
based on the subexpressions that appear in the predicate and
next-state expressions.  Details of the quantifier instantiation heuristic
can be found in an earlier work~\cite{lahiri-fmcad02}.
These subexpressions can contain symbols in
$\statesymbolset$, $\indexsymbolset$, and $\inputsymbolset$.  These
instantiated versions of the formulas enable the verifier to detect
specific cases where the predicates can be applied.  

More precisely, let $\subs$ be a substitution
assigning an expression $\subs_{\makesubsymbol{x}} \in
\exprset{\statesymbolset \cup \indexsymbolset \cup \inputsymbolset}$ for each
$\makesymbol{x} \in \indexsymbolset$.  
Then $\substituteone{\phi_{\makesubsymbol{p}}}{\subs}{\indexsymbolset}$ 
will be a Boolean expression over symbols
$\statesymbolset$, $\indexsymbolset$, and $\inputsymbolset$ that represents 
some instantiation of predicate $\phi_{\makesubsymbol{p}}$.

For a set of substitutions $\subsset$ and interpretations
$\interp{\indexsymbolset} \in \interpset{\indexsymbolset}$
and $\interp{\inputsymbolset} \in \interpset{\inputsymbolset}$,
we define the concretization function
$\gamma_{\subsset}$ as:
\begin{equation}
\gamma_{\subsset}(\absstateset, \interp{\indexsymbolset}, \interp{\inputsymbolset})
 \;\;\; \doteq \;\;\;
\left\{ s
 | \forall \subs \in \subsset :
\predevalsubs{s}{\interp{\indexsymbolset}}{\interp{\inputsymbolset}} \in \absstateset
\right \}
\label{eqn:gamma-approx}
\end{equation}

\begin{proposition}
For any abstract state set $\absstateset$ and interpretations $\interp{\indexsymbolset} \in \interpset{\indexsymbolset}$ and $\interp{\inputsymbolset} \in \interpset{\inputsymbolset}$:
\begin{enumerate}
\item
$\gamma(\absstateset) \subseteq
\gamma_{\subsset}(\absstateset, \interp{\indexsymbolset}, \interp{\inputsymbolset})$ for any set of substitutions $\subsset$.
\item
$\gamma_{\subsset}(\absstateset, \interp{\indexsymbolset}, \interp{\inputsymbolset}) \subseteq
\gamma_{\subssetp}(\absstateset, \interp{\indexsymbolset}, \interp{\inputsymbolset})$ for any pair of substitution sets $\subsset$
and $\subssetp$
satisfying $\subsset \supseteq \subssetp$.
\item For any abstract state set $\altabsstateset$, if 
$\absstateset \subseteq \altabsstateset$, then 
$\gamma_{\subsset}(\absstateset, \interp{\indexsymbolset}, \interp{\inputsymbolset}) \subseteq 
\gamma_{\subsset}(\altabsstateset, \interp{\indexsymbolset}, \interp{\inputsymbolset})$, for any set of substitutions $\subsset$.

\end{enumerate}
\label{prop:gamma-approx}
\end{proposition}

These properties follow directly from the definitions of $\gamma$ and
$\gamma_{\subsset}$ and Proposition 
\ref{prop:eval-sub}. 

\begin{proposition}
For any concrete state set $\concstateset$,
set of substitutions $\subsset$, and interpretations $\interp{\indexsymbolset} \in \interpset{\indexsymbolset}$ and $\interp{\inputsymbolset} \in \interpset{\inputsymbolset}$:
\begin{equation}
\concstateset \quad \subseteq \quad
\gamma_{\subsset}(\alpha(\concstateset), \interp{\indexsymbolset}, \interp{\inputsymbolset}).
\label{eqn:galois-gamma-approx}
\end{equation}
\label{prop:partial-galois-approx}
\end{proposition}

This property follows directly from Theorem \ref{thm:galois} and
Proposition \ref{prop:gamma-approx}.  It shows that for a given
interpretation $\interp{\indexsymbolset}$ and 
$\interp{\inputsymbolset}$, the functions $(\alpha,
\gamma_{\subsset})$ satisfy one of the properties of a Galois
connection (Equation~\ref{eqn:galois-gamma}), but they need not satisfy 
the other (Equation~\ref{eqn:galois-alpha}).  For example, when
$\subsset = \emptyset$, the quantified condition of
(\ref{eqn:gamma-approx}) becomes vacuous, and hence
$\gamma_{\subsset}(\absstateset, \interp{\indexsymbolset}, \interp{\inputsymbolset}) = \interpset{\statesymbolset}$.

We can use $\gamma_{\subsset}$ as an approximation to $\gamma$ in
defining the behavior of the abstract system.  That is, define
$\absapproxnext$ over sets of abstract states as:
\begin{eqnarray}
\absapproxnext(\absstateset) & = &
\left\{
\evalexpr{\substituteone{\phi}{\nextfunction}{\statesymbolset}}{\combineinterps{\combineinterps{s}{\interp{\indexsymbolset}}}{\interp{\inputsymbolset}}}
| \interp{\indexsymbolset} \in \interpset{\indexsymbolset},
\interp{\inputsymbolset} \in \interpset{\inputsymbolset},
s \in \gamma_{\subsset}(\absstateset, \interp{\indexsymbolset}, \interp{\inputsymbolset})
\right\}
\label{eqn:absapproxnext}
\\
& = &
\bigcup_{\interp{\indexsymbolset} \in \interpset{\indexsymbolset}}
\bigcup_{\interp{\inputsymbolset} \in \interpset{\inputsymbolset}}
\bigcup_{s \in \gamma_{\subs}(\absstateset, \interp{\indexsymbolset}, \interp{\inputsymbolset})}
\left\{\evalexpr{\substituteone{\phi}{\nextfunction}{\statesymbolset}}{\combineinterps{\combineinterps{s}{\interp{\indexsymbolset}}}{\interp{\inputsymbolset}}}\right\}
\label{eqn:absapproxnext-alt}
\end{eqnarray}
Observe in this equation that
$\substituteone{\phi_{\makesubsymbol{p}}}{\nextfunction}{\statesymbolset}$ is an expression describing the evaluation of predicate
$\phi_{\makesubsymbol{p}}$ in the next state.

It can be seen that $\absapproxnext(\absstateset) \supseteq
\absnext(\absstateset)$ for any set of abstract states $\absstateset$.
%
As long as $\subsset$ is nonempty (required to guarantee that $\absapproxnext$ is null-preserving), it can be shown that the
system defined by $\absapproxnext$ is an abstract
interpretation of the concrete system:
\begin{enumerate}
\item $\absapproxnext(\emptyset) = \emptyset$, if 
$\subsset$ is nonempty.
\item $\absapproxnext$ is monotonic: This follows from the definition
of $\absapproxnext$ in (\ref{eqn:absapproxnext-alt}) and 
Proposition~\ref{prop:gamma-approx}.
\item $\alpha(\concnext(\concstateset)) \subseteq 
\absapproxnext(\alpha(\concstateset))$: This follows from the fact
that $\alpha(\concnext(\concstateset)) \subseteq 
\absnext(\alpha(\concstateset))$ and $\absnext(\absstateset) 
\subseteq \absapproxnext(\absstateset)$. 
\end{enumerate}

We can therefore perform reachability analysis:
\begin{eqnarray}
\absreachapprox^{0} & = & \absinit  \label{eqn:approx-abs-reach-init}
\\
\displaystyle
\absreachapprox^{i+1} & = & 
\displaystyle
\absreachapprox^{i} \cup 
\absapproxnext(\absreachapprox^{i})
\label{eqn:approx-abs-reach-next}
\end{eqnarray}

These iterations will converge to a set $\absreachapprox$.  

\begin{proposition}
\mbox{}
\begin{enumerate}
\item
$\absreach \subseteq 
\absreachapprox$ for any set of substitutions $\subsset$.
\item
$\absreachapprox \subseteq
\absreachapproxp$ for any pair of substitution sets $\subsset$
and $\subssetp$ satisfying $\subsset \supseteq \subssetp$.
\end{enumerate}
\label{prop:reach-approx}
\end{proposition}

To see the first property, consider the following way of expressing the equation for $\absreach^{i+1}$
 (\ref{eqn:abs-reach-next}) using the alternative equation for $\alpha$ (\ref{eqn:alt-set-alpha}), and rearranging the order of the union operations:
\begin{eqnarray*}
\absreach^{i+1} & = & 
\displaystyle \absreach^{i} \;\cup 
\bigcup_{\interp{\indexsymbolset} \in \interpset{\indexsymbolset}} \;\;
\bigcup_{\interp{\inputsymbolset} \in \interpset{\inputsymbolset}} \;\;
\bigcup_{s \in \gamma(\absreach^{i})} \;\;
\left\{\evalexpr{\substituteone{\phi}{\nextfunction}{\statesymbolset}}{\combineinterps{\combineinterps{s}{\interp{\indexsymbolset}}}{\interp{\inputsymbolset}}}\right\}
\end{eqnarray*}
The property then follows by Proposition \ref{prop:gamma-approx}, using 
induction on $i$.  The second property also follows by
 Proposition \ref{prop:gamma-approx} using induction on $i$.

\begin{theorem}
For a formula $\psi \in \exprset{\predsymbolset}$, if 
$\evalexpr{\psi}{\interp{\predsymbolset}} = \true$
for every $\interp{\predsymbolset} \in \absreachapprox$,
then property 
$\forall \indexsymbolset \predmap{\psi}$ holds for the concrete system.
\label{thm:approx-safety}
\end{theorem}

\begin{proof}
Since $\evalexpr{\psi}{\interp{\predsymbolset}} = \true$
for every $\interp{\predsymbolset} \in \absreachapprox$ and 
$\absreach \subseteq \absreachapprox$ 
(by Proposition \ref{prop:reach-approx}), 
$\evalexpr{\psi}{\interp{\predsymbolset}} = \true$ for every
$\interp{\predsymbolset} \in \absreach$. Hence by 
Theorem \ref{thm:safety}, the property
$\forall \indexsymbolset \predmap{\psi}$ holds for the concrete system.

\end{proof}
This demonstrates that using quantifier instantiation during
reachability analysis yields a sound verification technique. However,
when the tool fails to verify a property, it could mean, in addition
to the two possibilities listed earlier, that 3) it used an inadequate
set of instantiations, or 4) that the property cannot be proved by any
bounded quantifier instantiation.

\section{Symbolic Formulation of Reachability Analysis}

We are now ready to express the reachability computation symbolically,
where each step involves finding the set of satisfying solutions to a
quantified CLU formula.
We will then see how this can be converted
into a problem of finding satisfying solutions to a Boolean formula.

On each step, we generate a Boolean formula $\absformapprox^i$, that
characterizes $\absreachapprox^i$.  That is
$\evalset{\absformapprox^i} = \absreachapprox^i$.  The formulas
directly encode the approximate reachability computations of
(\ref{eqn:approx-abs-reach-init}) and
(\ref{eqn:approx-abs-reach-next}).

Observe that by composing the predicate expressions with the initial
state expressions,
$\substituteone{\phi}{\indexstate{0}}{\statesymbolset}$, we get a set
of predicates over the 
initial state symbols $\initsymbolset$ indicating the conditions
under which the predicates hold in the initial state.  We can
therefore start the reachability analysis by finding solutions to the formula
\begin{eqnarray}
\absformapprox^0(\predsymbolset) & = &
\displaystyle{\exists \indexsymbolset \exists \initsymbolset 
 \bigwedge_{\makesubsymbol{p} \in \predsymbolset}
\makesymbol{p} \Leftrightarrow \substituteone{\phi}{\indexstate{0}}{\statesymbolset}}
\label{eqn:symbolic-reach-init}
\end{eqnarray}

\begin{proposition}
$\evalexpr{\absformapprox^0}{} = \absinit$
\end{proposition}

Let us understand the expression $\absformapprox^0$ by showing why it 
represents $\absinit$. Expanding the definition 
of $\absinit$, we get: 
\begin{equation}
\absinit = 
\bigcup_{\interp{\indexsymbolset} \in \interpset{\indexsymbolset}}
\bigcup_{s \in \concinit} 
\left\{\predeval{s}{\interp{\indexsymbolset}}\right\}
\label{eqn:repr-init1}
\end{equation}
Again, $\concinit = \bigcup_{\interp{\initsymbolset} \in \interpset{\initsymbolset}} \left\{\evalexpr{\indexstate{0}}{\interp{\initsymbolset}}\right\}$.
Using Proposition~\ref{prop:eval-sub}, we can rewrite (\ref{eqn:repr-init1})
as:
\begin{equation}
\absinit = 
\bigcup_{\interp{\indexsymbolset} \in \interpset{\indexsymbolset}}
\bigcup_{\interp{\initsymbolset} \in \interpset{\initsymbolset}}
\left\{\evalexpr{\substituteone{\phi}{\indexstate{0}}{\statesymbolset}}
{\combineinterps{\interp{\initsymbolset}}{\interp{\indexsymbolset}}}\right\}
\label{eqn:repr-init2}
\end{equation}

To generate a formula for the next-state computation, we 
first generate a formula for
$\gamma_{\subs}(\absreachapprox^i, \interp{\indexsymbolset}, \interp{\inputsymbolset})$ by forming a conjunction over each substitution in 
$\subsset$, where we compose the
current-state formula with the predicate expressions and with each
substitution $\subs$: $\bigwedge_{\subs \in \subsset}
\substituteone{\left(\substituteone{\absformapprox^i}{\phi}{\predsymbolset}\right)}{\subs}{\indexsymbolset}$. 


The formula for the next-state computation combines the alternate 
definition of $\absapproxnext$ (\ref{eqn:absapproxnext-alt}) and the 
formula for $\gamma_{\subsset}$ above:
\begin{eqnarray}
\absformapprox^{i+1}(\predsymbolset) & = &
\displaystyle{
\absformapprox^{i}(\predsymbolset) \; \lor \;
}
\nonumber \\
&& 
\exists \statesymbolset \exists \indexsymbolset \exists \inputsymbolset 
\displaystyle{
\left(
\bigwedge_{\subs \in \subsset}
\substituteone{\left(\substituteone{\absformapprox^i}{\phi}{\predsymbolset}\right)}{\subs}{\indexsymbolset}
 \;\; \land \;\;
\bigwedge_{\makesubsymbol{p} \in \predsymbolset} \makesymbol{p} \iff \substituteone{\phi_{\makesubsymbol{p}}}{\nextfunction}{\statesymbolset}
\right)}.
\label{eqn:symbolic-reach-next}
\end{eqnarray}

To understand the quantified term in this equation, note that the left-hand term is the formula for
$\gamma_{\subsset}(\absformapprox^i, \interp{\indexsymbolset},
\interp{\inputsymbolset})$, while the right-hand term expresses the conditions under which each abstract state variable $\makesymbol{p}$ will match the value of the corresponding predicate in the next state.


\begin{proposition}
$\evalexpr{\absformapprox^{i+1}}{} = \absreachapprox^{i+1}$
\end{proposition}

Let us see how this symbolic formulation would perform reachability
analysis for our example system.  
Recall that our system has two predicates 
$\phi_{\makesubsymbol{p}} \doteq
\makesymbol{F}(\makesymbol{x}) \geq 0$ and
$\phi_{\makesubsymbol{q}} \doteq
\makesymbol{x} \geq 0$.  In the initial state, $\makesymbol{F}$ is the function $\lambdaexpr{u}{u}$, and therefore 
$\substituteone{\phi_{\makesubsymbol{p}}}{\indexstate{0}}{\statesymbolset}$
simply becomes $\makesymbol{x} \geq 0$.  Equation
(\ref{eqn:symbolic-reach-init}) then becomes $\exists \makesymbol{x}
\left[
(\makesymbol{p} \iff \makesymbol{x} \geq 0) \land
(\makesymbol{q} \iff \makesymbol{x} \geq 0)
\right]$, which reduces to 
$\makesymbol{p} \iff \makesymbol{q}$.

Now let us perform the first iteration. 
For our instantiations we require two substitutions $\subs$ and $\subs'$ with
$\subs_{\makesubsymbol{x}} =
\makesymbol{x}$
and 
$\subs'_{\makesubsymbol{x}} = \incr{\makesymbol{i}}$.
For $\absformapprox^{0}(\makesymbol{p}, \makesymbol{q}) = 
\makesymbol{p} \iff \makesymbol{q}$,
the left-hand term of (\ref{eqn:symbolic-reach-next}) instantiates to
$(\makesymbol{F}(\makesymbol{x}) \geq 0 \iff \makesymbol{x} \geq 0)
\land (\makesymbol{F}(\incr{\makesymbol{i}}) \geq 0 \iff \incr{\makesymbol{i}} \geq 0)
$.
Substituting 
$\lambdaexpr{u}{\ITEe{u = \makesymbol{i}}{\makesymbol{F}(\incr{\makesymbol{i}})}{\makesymbol{F}(u)}}$
 for $\makesymbol{F}$ in
$\phi_{\makesubsymbol{p}}$ gives 
$(\compare{\makesymbol{x}}{\makesymbol{i}} \land
\makesymbol{F}(\incr{\makesymbol{i}}) \geq 0) \lor
(\ncompare{\makesymbol{x}}{\makesymbol{i}} \land
\makesymbol{F}(\makesymbol{x}) \geq 0)$.  

The quantified portion of (\ref{eqn:symbolic-reach-next}) for $\absformapprox^{1}(\makesymbol{p}, \makesymbol{q})$ then becomes:
\begin{equation}
\exists \; \makesymbol{F}, \makesymbol{x}, \makesymbol{i} :
\left (
\begin{array}{ll}
& 
\makesymbol{F}(\makesymbol{x}) \geq 0 \iff \makesymbol{x} \geq 0 
\;\;\land \;\;
\makesymbol{F}(\incr{\makesymbol{i}}) \geq 0 \iff \incr{\makesymbol{i}} \geq 0\\
\land &
\makesymbol{p} \iff [(\compare{\makesymbol{x}}{\makesymbol{i}} \land
\makesymbol{F}(\incr{\makesymbol{i}}) \geq 0) \lor
(\ncompare{\makesymbol{x}}{\makesymbol{i}} \land
\makesymbol{F}(\makesymbol{x}) \geq 0)] \\
\land &
\makesymbol{q} \iff \makesymbol{x} \geq 0
\end{array}
\right )
\label{eqn:example-reach}
\end{equation}
The only values of $\makesymbol{p}$ and $\makesymbol{q}$
 where this formula
cannot be satisfied is when $\makesymbol{p}$ is false and
$\makesymbol{q}$ is true.

As shown in \cite{lahiri-cav03a}, we can generate the set of solutions
to (\ref{eqn:symbolic-reach-init}) and (\ref{eqn:symbolic-reach-next})
by first transforming the formulas into equivalent Boolean formulas
and then performing quantifier elimination to remove all Boolean
variables other than those in $\predsymbolset$.  This quantifier
elimination is similar to the relational product operation used in
symbolic model checking and can be solved using either BDD or
SAT-based methods.

\section{Using a SAT Solver to Perform Reachability Analysis}

Observe that (\ref{eqn:symbolic-reach-next}) has a general form
$\chi'(\predsymbolset) = 
\chi(\predsymbolset) \lor
\exists \symbolseta \,
\theta(\symbolseta, \predsymbolset)$, where $\theta$ is a quantifier-free
CLU formula, $\symbolseta$ contains Boolean, integer,
function, and predicate symbols, and $\predsymbolset$ contains only
Boolean symbols.  Several methods (including those in~\cite{bryant-cav02,strichman-cav02,bryant-cfv02}) have been developed to transform a
quantifier-free CLU formula $\theta(\symbolseta, \predsymbolset)$ into a Boolean formula
$\tobool{\theta}(\tobool{\symbolseta}, \predsymbolset)$, where
$\tobool{\symbolseta}$ is now a set of Boolean variables,
in a way that preserves satisfiability.

By taking care~\cite{lahiri-cav03a}, this transformation can be performed 
in a way that preserves the set of satisfying solutions for the symbols in
$\predsymbolset$.  That is:
\begin{equation}
\{ \interp{\predsymbolset} | \exists \interp{\symbolseta} :
\evalexpr{\theta}{\combineinterps{\interp{\symbolseta}}{\interp{\predsymbolset}}} = \true \}
\;\; = \;\;
\{ \interp{\predsymbolset} | \exists \interp{\tobool{\symbolseta}} :
\evalexpr{\tobool{\theta}}{\combineinterps{\interp{\tobool{\symbolseta}}}{\interp{\predsymbolset}}} = \true \}
\label{eqn:clu-bool-trans}
\end{equation}
Based on such a transformation, we can generate a Boolean formula for $\chi'$ 
by repeatedly calling a Boolean SAT solver, yielding
one solution with each call.  In this presentation, we consider an
interpretation $\interp{\predsymbolset}$ to represent a Boolean
formula consisting of a conjunction of literals: $\makesymbol{p}$ when
$\interp{\predsymbolset}(\makesymbol{p}) = \true$ and
$\neg\makesymbol{p}$ when $\interp{\predsymbolset}(\makesymbol{p}) =
\false$.  Starting with $\chi' = \chi$,
and $\tobool{\theta}' = \tobool{\theta} \land \neg \chi$, we perform iterations:
\begin{eqnarray*}
\interp{\symbolseta}, \interp{\predsymbolset} & \leftarrow &
{\it SATSolve}(\tobool{\theta}')\\
\chi' &  \leftarrow & \chi' \lor \interp{\predsymbolset} \\
\tobool{\theta}' &  \leftarrow & \tobool{\theta}' \land \neg\interp{\predsymbolset} \\
\end{eqnarray*}
until $\tobool{\theta}'$ is unsatisfiable.

To illustrate this process, let us solve (\ref{eqn:example-reach}) to
perform the first iteration of reachability analysis on our example
system.  We can translate the right-hand term into Boolean form by
introducing Boolean variables $\makesymbol{a}$, $\makesymbol{b}$,
$\makesymbol{c}$, $\makesymbol{d}$ and $\makesymbol{e}$ encoding the predicates
$\makesymbol{F}(\makesymbol{x}) \geq 0$,
$\makesymbol{x} \geq 0$, 
$\makesymbol{F}(\incr{\makesymbol{i}}) \geq 0$,
$\makesymbol{\incr{\makesymbol{i}}} \geq 0$, 
and $\makesymbol{x} = \makesymbol{i}$, respectively.

The portion of
(\ref{eqn:example-reach}) within square brackets then becomes
$$\makesymbol{a} \Leftrightarrow \makesymbol{b} \land
\makesymbol{c} \Leftrightarrow \makesymbol{d} 
\land (\makesymbol{p} \Leftrightarrow 
\left[(\makesymbol{e} \land \makesymbol{c}) \lor 
(\neg \makesymbol{e} \land \makesymbol{a})
\right])
\land (\makesymbol{q} \Leftrightarrow \makesymbol{b}).
$$  
To this, let us add the consistency constraint: $\makesymbol{e} \land
\makesymbol{b} \Rightarrow \makesymbol{d}$ (encoding the property that
$\compare{\makesymbol{x}}{\makesymbol{i}} \land 
\makesymbol{x} \geq 0 \Rightarrow \incr{\makesymbol{i}} \geq
0$). Although the translation schemes will add a lot more constraints
(e.g., those involving uninterpreted function symbol), the above constraint
is sufficient to preserve the property described in 
(\ref{eqn:clu-bool-trans}). For simplicity, we will not describe the 
other constraints that would be added by the algorithms 
in ~\cite{lahiri-cav03a}. Finally, all the symbols apart from 
$\makesymbol{p}$ and $\makesymbol{q}$ are existentially quantified out.

It is easy to verify that the equation above with the consistency 
constraint is unsatisfiable only for the assignment when $\makesymbol{p}$
is false and $\makesymbol{q}$ is true. 

\section{Axioms}

As a special class of predicates, we may have some that are to hold at
all times.  For example, we could have an axiom
$\makesymbol{f}(\makesymbol{x}) > 0$ to indicate that function
$\makesymbol{f}$ is always positive, or $\makesymbol{f}(\makesymbol{y},
\makesymbol{z}) = \makesymbol{f}(\makesymbol{z}, \makesymbol{y})$ to
indicate that $\makesymbol{f}$ is commutative.  Typically, we want
these predicates to be individually quantified, but we can ensure this
by defining each of them over a unique set of index symbols, as we
have done in the above examples.

We can add this feature to our analysis by identifying a subset
$\axiomsymbolset$ of the predicate symbols $\predsymbolset$ to be
axioms.  We then want to restrict the analysis to states where the
axiomatic predicates hold.  Let Let $\aabsstateset$ denote the set of
abstract states $\interp{\predsymbolset}$ where
$\interp{\predsymbolset}(\makesymbol{p}) = \true$ for every
$\makesymbol{p} \in \axiomsymbolset$.  Then we can apply this restriction by redefining $\alpha(s)$ (Equation \ref{eqn:single-alpha})
for concrete state $s$ to be:
\begin{eqnarray}
\alpha(s) 
& \doteq &
\left \{ \predeval{s}
{\interp{\indexsymbolset}} | 
\interp{\indexsymbolset} \in \interpset{\indexsymbolset} \right\} \cap \aabsstateset
\label{eqn:single-alpha-axiom}
\end{eqnarray}
and then using this definition in the extension of $\alpha$ to sets
(Equation \ref{eqn:set-alpha}), the formulation of the reachability
analysis (Equations \ref{eqn:abs-reach-init} and
\ref{eqn:abs-reach-next}), and the approximate reachability analysis
(Equations \ref{eqn:approx-abs-reach-init} and
\ref{eqn:approx-abs-reach-next}).

The symbolic formulation of the approximate reachability analysis then becomes:
\begin{eqnarray*}
\absformapprox^0(\predsymbolset) & = &
\displaystyle{\exists \indexsymbolset \exists \initsymbolset
\left( \bigwedge_{\makesubsymbol{p} \in \predsymbolset-\axiomsymbolset}
\makesymbol{p} \Leftrightarrow \substituteone{\phi}{\indexstate{0}}{\statesymbolset}
\;\; \land \;\;
\bigwedge_{\makesubsymbol{p} \in \axiomsymbolset}
\substituteone{\phi}{\indexstate{0}}{\statesymbolset}
\right)}
\label{eqn:symbolic-reach-axiom-init}\\
\absformapprox^{i+1}(\predsymbolset) & = &
\displaystyle{
\absformapprox^{i}(\predsymbolset) \; \lor \; 
} 
\nonumber \\
&&
\exists \statesymbolset \exists \indexsymbolset \exists \inputsymbolset
\displaystyle{
\left(
\bigwedge_{\subs \in \subsset}
\substituteone{\left(\substituteone{\absformapprox^i}{\phi}{\predsymbolset}\right)}{\subs}{\indexsymbolset}
 \;\; \land \;\;
\bigwedge_{\makesubsymbol{p} \in \predsymbolset-\axiomsymbolset} \makesymbol{p} \Leftrightarrow \substituteone{\phi_{\makesubsymbol{p}}}{\nextfunction}{\statesymbolset}
\;\; \land \;\;
\bigwedge_{\makesubsymbol{p} \in \axiomsymbolset} \substituteone{\phi_{\makesubsymbol{p}}}{\nextfunction}{\statesymbolset}
\right)}.
\label{eqn:symbolic-reach-axiom-next}
\end{eqnarray*}

\section{Applications}
We have integrated the method described in this paper into
\uclid{}~\cite{bryant-cav02}, a tool for modeling and 
verifying infinite-state systems.
We have used our predicate abstraction tool to verify safety properties of 
a variety of models and protocols.  Some of the more interesting ones include:
\begin{enumerate}
\item A microprocessor out-of-order execution unit with an
unbounded retirement buffer.  Prior verification of this unit required
manually generating 13 invariants~\cite{lahiri-fmcad02}. 
The verification did not require any auxiliary invariants from the user
and the proof script (which consists of the 13 simple predicates)
is more compact than other verification efforts of similar models
based on compositional model checking~\cite{mcmillan-cav98} or theorem proving 
methods~\cite{arons-vlsi99,hosabettu-cav99}.
\item A directory-based cache protocol with unbounded channels, 
devised by Steven German of IBM \cite{german-cache}, as discussed below.
\item Versions of Lamport's bakery algorithm \cite{lamport-74} that
allow arbitrary number of processes to be active at each step or allow
non-atomic reads and writes.
\item Selection sort algorithm for sorting an arbitrary large array. We prove
the property that upon termination, the algorithm produces an ordered array. 
\item A model of the Ad-hoc On-demand Distance Vector (AODV) routing 
protocol~\cite{aodv}. This model was obtained from an earlier 
work~\cite{das-fmcad02}, where the protocol was verified  using
quantified predicates. 
\item A crucial invariant (similar to the one proved in~\cite{arons-cav01}) 
for proving the mutual exclusion for the Peterson's~\cite{peterson-ipl81} 
mutual exclusion algorithm. 
\end{enumerate}


\subsection{Directory-based Cache Coherence Protocol}

For the directory-based German's cache-coherence protocol, an unbounded
number of clients  ($\makesymbol{cache}$), communicate with a central 
{\it home} process to gain {\it exclusive} or {\it shared} access to a memory 
line. The state of each $\makesymbol{cache}$ can be
\{{\em invalid}, {\em shared}, {\em exclusive}\}. 
The home maintains explicit representations of two lists of clients:
those sharing the cache line ($\makesymbol{sharer\_list}$) 
and those for 
which the home has sent an invalidation request but has not received an 
acknowledgment ($\makesymbol{invalidate\_list}$) --- this prevents 
sending duplicate invalidation messages. 

The client places requests \{{\em req\_shared, req\_exclusive}\} on a 
channel $\makesymbol{ch\_1}$ and the home grants
\{{\em grant\_shared, grant\_exclusive}\} on channel $\makesymbol{ch\_2}$. 
The home also sends invalidation messages {\em invalidate} along 
$\makesymbol{ch\_2}$. 
The home grants
exclusive access to a client only when there are no clients sharing a
line, i.e. 
$\forall \makesymbol{i}: {\tt sharer\_list} (\makesymbol{i}) = \false$. 
The home maintains variables for the current client
($\makesymbol{ current\_client}$) and the current request 
($\makesymbol{current\_command}$). 
It also maintains a bit $\makesymbol{exclusive\_granted}$ to
indicate that some client has exclusive access. 
The cache lines acknowledge invalidation requests with a 
{\em invalidate\_ack} along another channel $\makesymbol{ch\_3}$.
At each step an input $\makesymbol{cid}$ is generated to denote the
process that is chosen at that step. Details of the 
protocol operation with single-entry channels can be found in many
previous works including~\cite{pnueli-tacas01}. We will refer to 
this version as {\it german-cache}. 

Since the modeling language of \uclid{} does not permit explicit quantifiers
in the system, we model the check for the absence of any sharers
$\forall \makesymbol{i}: {\tt sharer\_list} (\makesymbol{i}) = \false$
alternately. We maintain a Boolean state variable $\makesymbol{empty\_hsl}$,
which assumes an arbitrary value at each step of operation. 
We then add an axiom to the system: $\makesymbol{empty\_hsl} \iff 
\forall \makesymbol{i}: {\tt sharer\_list} (\makesymbol{i}) = \false$
\footnote{Our current implementation only handles one direction of the axiom,
$\forall \makesymbol{i}: \makesymbol{empty\_hsl} \myimplies
 {\tt sharer\_list} (\makesymbol{i}) = \false$, which is sufficient to
ensure the safety property.}. 
The quantified test $\forall \makesymbol{i}: {\tt sharer\_list} (\makesymbol{i}) = \false$ in the model is replaced by $\makesymbol{empty\_hsl}$. 

In our version of the protocol, each $\makesymbol{cache}$ 
communicates to the home process through three directed 
unbounded FIFO channels, namely the channels $\makesymbol{ch\_1,ch\_2,
ch\_3}$.  Thus, there are an unbounded number of 
unbounded channels, three for each client\footnote{The extension was 
suggested by Steven German himself}. 
It can be shown that a client can generate an unbounded number
of requests before getting a response from the home. We refer to this
version of the protocol as {\it german-cache-fifo}.

\noindent {\bf Proving Cache Coherence } 
We first consider the version {\it german-cache} which has been widely 
used in many previous 
works~\cite{pnueli-tacas01,emerson-charme03a,baukus-vmcai02}
among others and then consider the extended system {\it german-cache-fifo}. 
In both cases, the cache coherence property to prove
is $\forall i, j : \makesymbol{cache}(i) = $ {\em exclusive} $\wedge 
i \not= j \myimplies \makesymbol{cache}(j) = ${\em invalid}. 
All the experiments are performed on an 2.1GHz Pentium machine running 
Linux with 1GB of RAM. 

\subsubsection{Invariant Generation for {\it german-cache}} 
For this version, we derived two inductive invariants, one which involves 
a single process index $i$ and other which involves two process indices 
$i$ and $j$.

For single index invariant, we needed to add an auxiliary variable 
$\makesymbol{last\_granted}$ which tracks the last variable which has 
been granted exclusive access~\cite{pnueli-tacas01}. 
The inductive invariant which implies the cache coherence property was
constructed using the following set of predicates:

$\predsymbolset \doteq \{$
$\makesymbol{empty\_hsl}$,
$\makesymbol{exclusive\_granted}$,
$\makesymbol{current\_command} =$ {\em req\_shared},
$\makesymbol{current\_command} =$ {\em req\_exclusive},
$i = \makesymbol{last\_granted}$,
$\makesymbol{invalidate\_list}(i)$,
$\makesymbol{sharer\_list}(i)$,
$\makesymbol{cache}(i) =$ {\em exclusive}, 
$\makesymbol{cache}(i) =$ {\em invalid},
$\makesymbol{ch\_2}(i) = $ {\em grant\_exclusive}, 
$\makesymbol{ch\_2}(i) = $ {\em grant\_shared},
$\makesymbol{ch\_2}(i) = $ {\em invalidate},
$\makesymbol{ch\_3}(i) = $ {\em invalidate\_ack}
$\}$.

These predicates naturally appear in the system description. 
First, the predicates $\makesymbol{empty\_hsl}$ and
$\makesymbol{exclusive\_granted}$ are Boolean state variables. 
Next, for each enumerated state variable 
$\makesymbol{x}$, with range \{{\it $e_1,\ldots,e_m$}\}, we add the 
predicates $\makesymbol{x} =$ {\it $e_1$}, 
$\ldots$, $\makesymbol{x} =$ {\it $e_{m-1}$}, leaving the redundant 
predicate $\makesymbol{x} =$ {\it $e_m$}.
This explains $\makesymbol{current\_command} =$ {\em req\_shared} and
$\makesymbol{current\_command} =$ {\em req\_exclusive}. 
Next, we consider the values of the function and predicate state 
variables at a particular index $i$. In this example, such state variables
are the $\makesymbol{sharer\_list}$, $\makesymbol{invalidate\_list}$,
$\makesymbol{cache}$, $\makesymbol{ch\_1}$, $\makesymbol{ch\_2}$ and
$\makesymbol{ch\_3}$. We did not need to add any predicate for the 
$\makesymbol{ch\_1}$ since the content of this channel does not affect
the correctness condition. 
Finally, the predicate $i = \makesymbol{last\_granted}$ was added for the
auxiliary state variable $\makesymbol{last\_granted}$. 

With this set of 13 indexed predicates, the abstract reachability 
computation converged after 9 iterations in 14 seconds. Most of the
time (about 8 seconds) was spent in eliminating quantifiers from the
formula in (\ref{eqn:symbolic-reach-init}) and 
(\ref{eqn:symbolic-reach-next}) using the SAT-based quantifier 
elimination method. 

For the dual index invariant, addition of the second index 
variable $j$ makes the process computationally more expensive. 
However, the verification does not require any auxiliary variable to
prove the correctness. The set of predicates used is:

$\predsymbolset \doteq \{$
$\makesymbol{cache}(i) = $ {\em exclusive},
$\makesymbol{cache}(j) = $ {\em invalid},
$i = j$,
$\makesymbol{ch2}(i) =$ {\em grant\_exclusive},
$\makesymbol{ch2}(i) =$ {\em grant\_shared},
$\makesymbol{ch2}(i) =$ {\em invalidate},
$\makesymbol{ch3}(i) =$ {\em empty},
$\makesymbol{ch2}(j) =$ {\em grant\_exclusive},
$\makesymbol{ch2}(j) =$ {\em grant\_shared},
$\makesymbol{ch2}(j) =$ {\em invalidate},
$\makesymbol{ch3}(j) =$ {\em empty},
$\makesymbol{invalidate\_list}(i)$,
$\makesymbol{current\_command} =$ {\em req\_exclusive},
$\makesymbol{current\_command} =$ {\em req\_shared},
$\makesymbol{exclusive\_granted}$,
$\makesymbol{sharer\_list}(i)$,
$\}$.

The inductive invariant which implies the cache-coherency was constructed
using these 16 predicates in 41 seconds using 12 steps of abstract 
reachability. The portion of time spent on eliminating quantifiers was around
15 seconds.

\subsubsection{Invariant Generation for {\it german-cache-fifo}}
For this version, each of the channels, namely $\makesymbol{ch1}$, 
$\makesymbol{ch2}$ and $\makesymbol{ch3}$ are modeled as unbounded FIFO
buffers. Each channel has a head (e.g. $\makesymbol{ch1\_hd}$), which is the position
of the earliest element in the queue and a tail pointer
(e.g. $\makesymbol{ch1\_tl}$), which is the position of the first {\it free}
entry for the queue, where the next element is inserted.
These pointers are modeled as function state variables, which maps process
$\makesymbol{i}$ to the value of the head or tail  pointer of a channel for that process. 
For instance, $\makesymbol{ch2\_hd}(\makesymbol{i})$ denotes the 
position of the head pointer for the process $\makesymbol{i}$. 
The channel itself is modeled as a two-dimensional array, where 
$\makesymbol{ch2}(\makesymbol{i},\makesymbol{j})$ denotes the content of the 
channel at index $\makesymbol{j}$ for the process $\makesymbol{i}$. 
We aim to derive an invariant over a single process index $i$ and an index
$j$ for an arbitrary element of the channels. Hence we 
add the auxiliary variable $\makesymbol{last\_granted}$. The set of 
predicates required for this model is:

$\predsymbolset \doteq \{$
$\makesymbol{cache} (i) = $ {\em exclusive},
$\makesymbol{cache} (i) = $ {\em invalid},
$i = \makesymbol{last\_granted}$,
$\makesymbol{current\_command} =$ {\em req\_shared},
$\makesymbol{current\_command} =$ {\em req\_exclusive},
$\makesymbol{exclusive\_granted}$,
$\makesymbol{invalidate\_list}(i)$,	
$\makesymbol{sharer\_list}(i)$,
$j = \makesymbol{ch2\_hd}(i)$,
$j = \makesymbol{ch3\_hd}(i)$,
$j \leq \makesymbol{ch2\_hd}(i)$,
$j < \makesymbol{ch2\_tl}(i)$,
$j \leq \makesymbol{ch3\_hd}(i)$,
$j < \makesymbol{ch3\_tl}(i)$,
$j = \decr{\makesymbol{ch2\_tl}(i)}$,
$\makesymbol{ch1\_hd}(i) < \makesymbol{ch1\_tl}(i)$,
$\makesymbol{ch1\_hd}(i) = \makesymbol{ch1\_tl}(i)$,
$\makesymbol{ch2\_hd}(i) < \makesymbol{ch2\_tl}(i)$,
$\makesymbol{ch2\_hd}(i) = \makesymbol{ch2\_tl}(i)$,
$\makesymbol{ch2}(i,j) =$ {\em grant\_exclusive},
$\makesymbol{ch2}(i,j) =$ {\em grant\_shared},
$\makesymbol{ch2}(i,j) =$ {\em invalidate},
$\makesymbol{ch3\_hd}(i) < \makesymbol{ch3\_tl}(i)$,
$\makesymbol{ch3\_hd}(i) = \makesymbol{ch3\_tl}(i)$,
$\makesymbol{ch3\_tl}(i) = \incr{\makesymbol{ch3\_hd}(i)}$,
$\makesymbol{ch3}(i,j) =$ {\em invalidate\_ack}
$\}$.

Apart from the predicates required for {\it german-cache}, we require
predicates involving entries in the various channels for a particular 
cache entry $i$. Predicates like
$\makesymbol{ch1\_hd}(i) < \makesymbol{ch1\_tl}(i)$ and 
$\makesymbol{ch1\_hd}(i) = \makesymbol{ch1\_tl}(i)$ are used to determine
if the particular channel is non-empty. To reason about {\it active} 
entries in a FIFO, i.e., those lying between the head (inclusive) and the 
tail, we need  predicates like $j \leq \makesymbol{ch2\_hd}(i)$ and
$j < \makesymbol{ch2\_tl}(i)$. The content of the channel at a location 
$j$ is given by the predicates like $\makesymbol{ch2}(i,j) =$ {\em grant\_exclusive}
and $\makesymbol{ch3}(i,j) =$ {\em invalidate\_ack}. Finally, a couple of 
predicates like $\makesymbol{ch3\_tl}(i) = \incr{\makesymbol{ch3\_hd}(i)}$ and
$j = \decr{\makesymbol{ch2\_tl}(i)}$ are added by looking at failures to prove
the cache coherence property.

Our tool constructs an inductive invariant with these 26 predicates
which implies the cache coherence property. The abstract reachability
took 17 iterations to converge in 1435 seconds. The quantifier elimination
process took 1227 seconds.


\section*{Acknowledgments}

We wish to thank Ching-Tsun Chou for his detailed comments on an early
draft of this paper.

\bibliography{refs}

\end{document}